\documentclass[preprint,amsmath,amssymb,aps,prd,a4paper]{revtex4-2}

\usepackage{xcolor}
\usepackage{graphicx}
\usepackage{amssymb}
\usepackage{multirow}
\usepackage{array}
\usepackage{braket}
\usepackage{hyperref}
\usepackage{diagbox}
\usepackage{slashed}

\newcommand{\bfk}{{\bf k}_{\perp}}

\usepackage{xcolor}
\begin{document}
\title{Parton distribution functions of the proton in a hot isospin-asymmetric nuclear medium and finite nuclei}

\author{Navpreet Kaur}
\email{knavpreet.hep@gmail.com}
\author{Manpreet
	Kaur}
\email{
	ranapreeti803@gmail.com}
\author{Dhananjay Singh}
\email{
	snaks16aug@gmail.com}
\author{Arvind Kumar}
\email{kumara@nitj.ac.in}
\author{Harleen Dahiya}
\email{dahiyah@nitj.ac.in}
\author{Suneel Dutt}
\email{dutts@nitj.ac.in}

\affiliation{Department of Physics, Dr. B.R. Ambedkar National Institute of Technology, Jalandhar, 144008, India}

\date{\today}
\begin{abstract}
    When a nucleon is bound in a nucleus, its quark distributions are modified, and the change in the spin-dependent distributions can differ in size from that in the spin-independent ones. We compute the unpolarized $f_1(x)$, helicity $g_1(x)$, and transversity $h_1(x)$ distributions of the $u$ and $d$ quarks in a proton immersed in hot and isospin-asymmetric nuclear matter. The proton is described in the diquark spectator model, with the medium effects entering through the effective quark, diquark, and proton masses computed using the chiral SU(3) quark mean field (CQMF) model. The axial and tensor charges are obtained for baryon densities up to saturation density and temperatures $T = 0$ and $0.1$ GeV, in both symmetric nuclear matter and pure neutron matter. The distributions are then carried to $^{40}$Ca, $^{56}$Fe, $^{197}$Au, and $^{208}$Pb at zero temperature through the local density approximation. In symmetric matter at saturation density and zero temperature, the peaks of the three $u$ quark distributions move toward larger $x$ and fall least for the unpolarized distribution and most for the transversity distribution, so the spin-dependent $u$ quark distributions are modified more strongly. Between vacuum and saturation density, the axial charge of the proton drops from $1.28$ to $0.844$, and the tensor charge of the $u$ quark decreases while that of the $d$ quark grows in magnitude. At the center of each nucleus, the axial charge of the proton rises from $0.699$ in $^{40}$Ca to $0.817$ in $^{208}$Pb as the central baryon density decreases, which suggests that the modification at the center follows the central density rather than the mass number.
\end{abstract}
%
\maketitle
\section{Introduction}
The internal structure of a nucleon bound in a nucleus differs from that of a free nucleon. The European Muon Collaboration (EMC) measured the ratio $F_2^N({\rm Fe})/F_2^N(D)$ of the structure functions per nucleon for iron and deuterium, and found the iron value to be smaller \cite{Aubert:1983xm}. Electron scattering at SLAC confirmed the effect on steel and aluminum \cite{Bodek:1983ec,Bodek:1983qn}, and a later measurement traced its atomic weight dependence over nine nuclei from deuterium to gold \cite{Arnold:1983mw}. Fermi motion alone predicts a cross-section ratio $\sigma_{\rm Fe}/\sigma_{D_2} = 1.069$ at a longitudinal momentum fraction $x = 0.65$, where the measurement gives $0.870 \pm 0.017$ \cite{Bodek:1983ec}. The effect does not correlate directly with the binding energy per nucleon, which peaks near iron while the cross-section ratios continue to fall for heavier nuclei \cite{Arnold:1983mw}. This suggests that the binding energy alone may not govern the size of the effect. Corrections applied at the nuclear level, through binding and Fermi motion, are not sufficient, and reproducing the data requires the internal quark structure of the bound nucleon itself to be modified \cite{Cloet:2006bq}. Interpretation in a finite nucleus is complicated because even in $^{197}$Au, fewer than half of the nucleons sit in the constant-density interior, and the experimental data have therefore been extrapolated to infinite nuclear matter in the local density approximation \cite{Sick:1992pw}. In the present work, the proton is placed in such a medium, and we compute its distributions as the baryon density, temperature, and isospin asymmetry are varied separately.  The distributions are then carried to the finite nuclei $^{40}$Ca, $^{56}$Fe, $^{197}$Au, and $^{208}$Pb at zero temperature through the local density approximation. For the proton and neutron densities, we use Woods-Saxon distributions with the radius and diffuseness parameters of Ref. \cite{Chamon:2002}, obtained from fits to Dirac-Hartree-Bogoliubov densities, as in Ref. \cite{Ma:2024}.

The quark structure of the proton at leading twist is described by three distributions, the unpolarized $f_1(x)$, the helicity $g_1(x)$, and the transversity $h_1(x)$. The distribution $f_1$ counts the quarks of a given flavor. The other two give the difference between quarks polarized parallel and antiparallel to the proton spin, longitudinally for $g_1$ and transversely for $h_1$. In Ref. \cite{Jaffe:1991kp}, the light-front quark correlation function is shown to require nine distribution functions, three each at twists 2, 3, and 4. Only $f_1$, $g_1$, and $h_1$ enter a complete quark-parton model of the nucleon at leading twist. Of the three, only $h_1$ is chiral odd. It does not contribute to inclusive deep inelastic scattering (DIS), where $f_1$ and $g_1$ are measured \cite{Jaffe:1991kp}. The distribution $h_1$ originates in the analysis of dimuon production from polarized proton-proton collisions, where the average transverse spin of the quarks enters the cross section alongside their average helicity \cite{Ralston:1979ys}. 
The first extraction of $h_1$ within collinear factorization used pion-pair production in semi-inclusive DIS and in proton-proton collisions, where $h_1$ pairs with a dihadron fragmentation function in a simple product \cite{Radici:2018iag}. A recent global analysis of dihadron production extracted $h_1$ together with the fragmentation functions, and determined the tensor charge $g_T^q$, its first moment, both with and without lattice results included as a prior \cite{Cocuzza:2024}. In contrast, the helicity distributions and their uncertainties have been determined from a global analysis of polarized DIS and proton-proton data \cite{deFlorian:2008mr}. These results show how unevenly the three distributions are known in vacuum, with transversity the least constrained of the three.

In-medium nucleon distributions have been computed in several quark-level frameworks. In Ref. \cite{Mineo:2003vc}, the Nambu-Jona-Lasinio (NJL) model is used to treat the nucleon as a relativistic quark-diquark state in mean-field nuclear matter. The quenching in the valence region was found to come from the vector field, while the scalar potential, taken together with Fermi motion, produces an enhancement near $x \simeq 0.7$. A further NJL calculation obtains the nucleon from the homogeneous Faddeev equation with proper-time regularization, and the constituent quark mass falls from 0.40 to 0.32 GeV and the nucleon mass from 0.94 to 0.75 GeV at saturation density \cite{Cloet:2005rt}. Carried to finite nuclei through a shell model and a convolution, it found the spin-independent distributions quenched at large $x$ and enhanced at small $x$, and the polarized effect roughly twice the unpolarized one \cite{Cloet:2006bq}. The quark-meson coupling (QMC) model supplies the medium to a nucleon built in the MIT bag, where the dominant contributions are the $\sigma$-generated masses and the $\omega$ vector potential, the polarized and unpolarized effects being about the same size with an axial-charge suppression of $8.9\%$ \cite{Tronchin:2018mvu}. The chiral quark-soliton model reaches nuclear-matter saturation from an attractive nucleon-nucleon interaction with vector-meson exchange, and describes the EMC effect together with the related Drell-Yan measurements \cite{Smith:2003hu}. In a light-front constituent quark treatment with QMC input, the quark momentum is shifted by the vector potential alongside the replacement of the masses \cite{deAraujo:2017uad}. The present calculation carries no vector potential, and the medium enters through the effective quark, diquark, and proton masses alone, which in the finite nuclei are evaluated at the local baryon density and isospin asymmetry. We therefore compute the response of the distributions to those masses. Since the valence quenching in Ref. \cite{Mineo:2003vc} comes from the vector field, the distributions obtained here may not show the quenching seen in the EMC measurements.

All three distributions of the free proton have been obtained in the valence sector of Basis Light-Front Quantization (BLFQ), as forward limits of the corresponding generalized parton distributions (GPDs) \cite{SKaur:2024}. With one dynamical gluon in the Fock space, the same approach gives the unpolarized and helicity quark distributions together with the gluon distribution \cite{Xu:2023}. That calculation has since been embedded in QMC nuclear matter, where the valence component of the proton loses probability to the quark-gluon component as the density rises \cite{Wu:2026mps}. That in-medium calculation covers unpolarized distributions in symmetric matter at zero temperature. In Ref. \cite{Cloet:2009}, the NJL quark-diquark treatment is extended to asymmetric matter with an isovector-vector $\rho^0$ field, and the EMC effect is found to be largest near $Z/N \simeq 0.6$. We are not aware of an in-medium calculation of the proton transversity distribution. None of the in-medium calculations cited above is carried out at finite temperature.

For the mesons, in-medium valence distributions of the pion have been obtained from a light-front wave function (LFWF) carrying the QMC effective quark masses \cite{deMelo:2016uwj}. The regulator mass is kept at its vacuum value of 600 MeV, and the kaon calculation makes the same choice \cite{Yabusaki:2023zin}. In the light-front quark model, the Gaussian parameter has been held density independent because its medium dependence is not well understood \cite{Arifi:2023jfe}, and fixing the harmonic parameter leaves the constituent mass from the NJL model as the only medium-dependent input \cite{Dwibedi:2025vhr}. In Ref. \cite{Hutauruk:2019ipp}, the NJL model with QMC input gives the pion and kaon valence distributions with light-quark momenta shifted by the vector potential and rescaled $x$, subject to the baryon number and momentum sum rules in the medium \cite{Hutauruk:2021kej}. All these calculations stay in symmetric nuclear matter at zero temperature. 

Valence quark distributions of the pion have been obtained in isospin-asymmetric nuclear matter at finite temperature, with the effective quark masses taken from the chiral SU(3) quark mean field (CQMF) model \cite{Puhan:2024xdq}. The distributions respond more strongly to baryon density than to temperature or isospin asymmetry, and the peak of the valence distribution moves toward larger $x$ with increasing density. The same construction has been extended to the kaon and antikaon doublets, whose valence quark and antiquark carry unequal masses \cite{Singh:2024lra}. In a denser medium, the light quark carries a smaller fraction of the longitudinal momentum and the strange antiquark a larger one. In the CQMF model, quarks are confined inside the baryon and interact through the scalar fields $\sigma$ and $\zeta$ together with the vector fields $\omega$ and $\rho$ \cite{Wang:2001CTP}. The broken scale invariance of QCD enters through a dilaton field $\chi$ that reproduces the trace anomaly \cite{Papazoglou:1998vr}. The scalar-isovector field $\delta$ vanishes in symmetric matter and splits the effective masses of the $u$ and $d$ quarks when the medium is asymmetric \cite{Singh:2017mxj}. The pion and kaon calculations thus describe the valence quark structure of the pseudoscalar mesons at finite baryon density, temperature, and isospin asymmetry.

Motivated by this, we aim to examine the influence of the medium on the three leading-twist parton distributions of the proton within the diquark spectator model \cite{Bacchetta:2008af}, with the in-medium quark, diquark, and proton masses taken from the CQMF model. In vacuum, the same diquark model has been used to study the chiral-odd GPDs of the octet baryons, with the transversity distributions and tensor charges obtained in the forward limit \cite{Kaur:2025}. We evaluate the distributions of the $u$ and $d$ quarks across a range of baryon density, temperature, and isospin asymmetry, and obtain from them the second Mellin moment $\langle x \rangle^q$ of $f_1$, together with the axial charge $\Delta q$ and the tensor charge $\delta q$ as the first moments of $g_1$ and $h_1$. The same three distributions are then obtained at zero temperature for a proton at the center of each of the four nuclei, and as a function of the radial distance in $^{208}$Pb. Compared with the in-medium calculations cited above, we obtain in addition the transversity distribution and the tensor charge of the proton in the medium, and we follow all three distributions at finite temperature. Medium effects are carried by the effective masses alone, and we use the vacuum definition of the longitudinal momentum fraction, as was done for the pion and the kaon \cite{Puhan:2024xdq,Singh:2024lra}. 

The rest of this paper is organized as follows. In Sec. \ref{Sec2}, we outline the CQMF model and the in-medium masses it supplies. In Sec. \ref{sec:pdf}, we introduce the diquark spectator model and the leading-twist distributions, and discuss how they respond to the baryon density, temperature, and isospin asymmetry of the medium. In Sec. \ref{Sec:finite}, we obtain the distributions of a proton bound in finite nuclei. Finally, in Sec. \ref{summary}, we summarize our findings.

\section{Chiral SU(3) quark mean field model \label{Sec2}}
We obtain the in-medium quark and baryon masses from the CQMF model, in which the total effective Lagrangian density is written as \cite{Singh:2017mxj}
\begin{equation}
    \mathcal{L}_{\rm eff} = \mathcal{L}_{q0} + \mathcal{L}_{qm} + \mathcal{L}_{\Sigma\Sigma} + \mathcal{L}_{VV} + \mathcal{L}_{\chi SB} + \mathcal{L}_{\Delta m} + \mathcal{L}_{c},
    \label{eq:Leff}
\end{equation}
where $\mathcal{L}_{q0} = \bar{\psi}\, i\gamma^\mu \partial_\mu \psi$ is the free part for massless quarks, $\mathcal{L}_{qm}$ is the chiral SU(3) invariant quark-meson interaction, and $\mathcal{L}_{\Sigma\Sigma}$ and $\mathcal{L}_{VV}$ are the self-interactions of the scalar and vector mesons. The term $\mathcal{L}_{\chi SB}$ breaks chiral symmetry explicitly, $\mathcal{L}_{\Delta m} = -\Delta m_s\, \bar{\psi} S_1 \psi$ with $S_1 = {\rm diag}(0,0,1)$ shifts the mass of the strange quark, and $\mathcal{L}_c = -\bar{\psi} \chi_c \psi$ confines the quarks. In the mean-field approximation, the meson field operators are replaced by their classical expectation values, only the timelike components of the vector fields survive, and the pseudoscalar fields vanish by parity conservation \cite{Papazoglou:1998vr}. For nuclear matter, we retain the scalar fields $\sigma$, $\zeta$, and $\delta$, the dilaton $\chi$, and the vector fields $\omega$ and $\rho$. The couplings of the quarks to the scalar-isoscalar fields $\sigma$ and $\zeta$ follow from the SU(3) flavor structure of $\mathcal{L}_{qm}$ and are expressed through a single coupling constant $g_s$ \cite{Wang:2001CTP},
\begin{equation}
    g_\sigma^u = g_\sigma^d = \frac{g_s}{\sqrt{2}}, \quad g_\zeta^s = g_s, \quad g_\zeta^u = g_\zeta^d = g_\sigma^s = 0.
    \label{eq:gq}
\end{equation}
The effective quark masses are written as \cite{Singh:2024lra}
\begin{equation}
    m_q^* = -g_\sigma^q \sigma - g_\zeta^q \zeta - g_\delta^q I_3^q \delta + \Delta m_q,
    \label{eq:mq}
\end{equation}
where $I_3^q$ is the third component of isospin of the quark, with $I_3^u = -I_3^d = 1/2$ and $I_3^s = 0$, and $\Delta m_u = \Delta m_d = 0$. The $\delta$ field couples to the $u$ and $d$ quarks with the same strength, $g_\delta^u = g_\delta^d = g_s/\sqrt{2}$, and not to the $s$ quark $g_\delta^s = 0$, so the sign of its contribution to each quark mass is fixed by $I_3^q$ alone \cite{Singh:2024lra}. Equations (\ref{eq:gq}) and (\ref{eq:mq}) then give
\begin{equation}
    m_u^* = -\frac{g_s}{\sqrt{2}}\left(\sigma + \frac{\delta}{2}\right), \quad m_d^* = -\frac{g_s}{\sqrt{2}}\left(\sigma - \frac{\delta}{2}\right), \quad m_s^* = -g_s \zeta + \Delta m_s.
    \label{eq:mud}
\end{equation}
The vacuum values of the nonstrange and strange scalar fields are fixed by the pion and kaon decay constants as $\sigma_0 = -f_\pi$ and $\zeta_0 = (f_\pi - 2f_K)/\sqrt{2}$ \cite{Papazoglou:1998vr}.

The scalar meson self-interaction, including the scalar-isovector field $\delta$, is written as \cite{Papazoglou:1998vr,Singh:2017mxj}
\begin{align}
    \mathcal{L}_{\Sigma\Sigma} &= -\frac{1}{2}k_0\chi^2\left(\sigma^2+\zeta^2+\delta^2\right) + k_1\left(\sigma^2+\zeta^2+\delta^2\right)^2 + k_2\left(\frac{\sigma^4}{2}+\frac{\delta^4}{2}+3\sigma^2\delta^2+\zeta^4\right) \nonumber\\
    &\quad + k_3\chi\left(\sigma^2-\delta^2\right)\zeta - k_4\chi^4 - \frac{1}{4}\chi^4\ln\frac{\chi^4}{\chi_0^4} + \frac{\xi}{3}\chi^4\ln\frac{\left(\sigma^2-\delta^2\right)\zeta}{\sigma_0^2\zeta_0}.
    \label{eq:LSS}
\end{align}
The first logarithmic term, $-\frac{1}{4}\chi^4\ln(\chi^4/\chi_0^4)$, breaks scale invariance and makes the trace of the energy-momentum tensor proportional to $\chi^4$. Comparison with the trace anomaly of QCD then identifies the dilaton with the gluon condensate. The coefficient $\xi$ of the second logarithm is guided by the one-loop QCD $\beta$ function, which suggests $\xi = 6/33$ for three colors and three flavors \cite{Papazoglou:1998vr}. The vector meson self-interaction is written as
\begin{equation}
    \mathcal{L}_{VV} = \frac{1}{2}\frac{\chi^2}{\chi_0^2}\left(m_\omega^2\omega^2 + m_\rho^2\rho^2\right) + g_4\left(\omega^4 + 6\omega^2\rho^2 + \rho^4\right),
    \label{eq:LVV}
\end{equation}
where the masses $m_\omega$ and $m_\rho$ are taken as $783$ MeV. The explicit chiral symmetry breaking term, which gives the pseudoscalar mesons nonvanishing masses and satisfies the partially conserved axial-vector current relations for the pion and the kaon, is written as
\begin{equation}
    \mathcal{L}_{\chi SB} = -\frac{\chi^2}{\chi_0^2}\left[m_\pi^2 f_\pi \sigma + \left(\sqrt{2}\, m_K^2 f_K - \frac{m_\pi^2 f_\pi}{\sqrt{2}}\right)\zeta\right].
    \label{eq:LSB}
\end{equation}

The confining term $\mathcal{L}_c$ in Eq. (\ref{eq:Leff}) is built from the scalar-vector potential \cite{Wang:2001CTP,Singh:2017mxj}
\begin{equation}
    \chi_c(r) = \frac{1}{4}k_c r^2\left(1+\gamma^0\right),
    \label{eq:chic}
\end{equation}
and each quark in the meson mean fields obeys the Dirac equation
\begin{equation}
    \left[-i\boldsymbol{\alpha}\cdot\boldsymbol{\nabla} + \chi_c(r) + \beta m_q^*\right]\psi_q = e_q\,\psi_q,
    \label{eq:Dirac}
\end{equation}
where $\boldsymbol{\alpha}$ and $\beta$ are the Dirac matrices and $m_q^*$ is given by Eq. (\ref{eq:mq}). For this potential, the energy $e_q$ of a quark in the lowest ($1S$) state follows from the condition
\begin{equation}
    e_q = m_q^* + \frac{3\sqrt{k_c}}{\sqrt{2\left(e_q + m_q^*\right)}}.
    \label{eq:eigen}
\end{equation}
The strength of the confining potential, $k_c = 300$ MeV fm$^{-2}$, is fixed by requiring the model to reproduce the saturation properties of symmetric nuclear matter.

The energy of a nucleon built from three quarks of energy $e_q$ is written as \cite{Singh:2017mxj,Singh:2024lra}
\begin{equation}
    E_i = \sum_q n_q^i e_q + E_{\rm spin}^i,
    \label{eq:EN}
\end{equation}
where $n_q^i$ is the number of quarks of flavor $q$ in the nucleon $i = p, n$ and $E_{\rm spin}^i$ accounts for the spin correlations \cite{Shen:2000}. The independent-quark state is not translationally invariant, so $E_i$ contains the energy of a spurious center-of-mass motion \cite{Barik:1985}. The effective nucleon mass in the medium is obtained by removing this contribution \cite{Barik:1985,Shen:2000},
\begin{equation}
    M_i^* = \sqrt{E_i^2 - \langle P_{\rm c.m.}^2\rangle_i}, \qquad \langle P_{\rm c.m.}^2\rangle_i = \sum_q n_q^i \langle p_q^2\rangle,
    \label{eq:MN}
\end{equation}
where the quarks are taken to move independently in the confining potential. For the equally mixed scalar-vector harmonic potential of Eq. (\ref{eq:chic}), the mean square momentum of a quark in its $1S$ state is \cite{Barik:1985,Singh:2024lra}
\begin{equation}
    \langle p_q^2\rangle = \frac{\left(11e_q + m_q^*\right)\left(e_q^2 - m_q^{*2}\right)}{6\left(3e_q + m_q^*\right)}.
    \label{eq:p2}
\end{equation}
Following Ref. \cite{Shen:2000}, $E_{\rm spin}^i$ is fixed by the free nucleon mass and is assumed not to change in the nuclear medium. We take $E_{\rm spin}^p = E_{\rm spin}^n = -567.012$ MeV, which reproduces the nucleon mass in vacuum $M_N = 939.02$ MeV. The $\Delta$ baryons are not included as constituents of the medium. Their masses are evaluated in the same way from the quark energies in nuclear matter, with $E_{\rm spin}^\Delta$ fixed by the vacuum mass $M_\Delta = 1232$ MeV.

At finite temperature $T$, the thermodynamic potential per unit volume of isospin-asymmetric nuclear matter is written as \cite{Singh:2017mxj,Singh:2024lra}
\begin{equation}
    \Omega = -\frac{\gamma_N T}{(2\pi)^3}\sum_{i=p,n}\int d^3k\left\{\ln\left[1+e^{-\left(E_i^*(k)-\mu_i^*\right)/T}\right] + \ln\left[1+e^{-\left(E_i^*(k)+\mu_i^*\right)/T}\right]\right\} - \mathcal{L}_M - \mathcal{V}_{\rm vac},
    \label{eq:Omega}
\end{equation}
where $\gamma_N = 2$ is the spin degeneracy of the nucleons, $E_i^*(k) = \sqrt{k^2 + M_i^{*2}}$, $\mathcal{L}_M = \mathcal{L}_{\Sigma\Sigma} + \mathcal{L}_{VV} + \mathcal{L}_{\chi SB}$, and $\mathcal{V}_{\rm vac}$ is subtracted to obtain zero vacuum energy. The vector fields shift the chemical potential $\mu_i$ of the nucleon to the effective value \cite{Wang:2001CTP}
\begin{equation}
    \mu_i^* = \mu_i - g_\omega^N \omega - g_\rho^N I_3^i \rho,
    \label{eq:mustar}
\end{equation}
with $I_3^p = -I_3^n = 1/2$. The vector and scalar densities of the nucleon $i$ are
\begin{equation}
    \rho_v^i = \frac{\gamma_N}{(2\pi)^3}\int d^3k\left[f_i(k) - \bar{f}_i(k)\right], \qquad \rho_s^i = \frac{\gamma_N}{(2\pi)^3}\int d^3k\,\frac{M_i^*}{E_i^*(k)}\left[f_i(k) + \bar{f}_i(k)\right],
    \label{eq:rho}
\end{equation}
where $f_i(k) = \{1 + e^{[E_i^*(k)-\mu_i^*]/T}\}^{-1}$ and $\bar{f}_i(k) = \{1 + e^{[E_i^*(k)+\mu_i^*]/T}\}^{-1}$ are the particle and antiparticle distribution functions. The baryon density is $\rho_B = \rho_v^p + \rho_v^n$, and the isospin asymmetry of the medium is described by $\eta = (\rho_v^n - \rho_v^p)/(2\rho_B)$, which vanishes in symmetric nuclear matter and equals $0.5$ in pure neutron matter. The meson fields follow from the stationarity conditions $\partial\Omega/\partial\varphi = 0$, where $\varphi$ stands for each of $\sigma$, $\zeta$, $\delta$, $\chi$, $\omega$, and $\rho$. The nucleon term of Eq. (\ref{eq:Omega}), which we denote by $\Omega_N$, depends on $\sigma$ and $\delta$ only through $M_i^*$ and on $\omega$ and $\rho$ only through $\mu_i^*$. At fixed $T$ and $\mu_i$, its derivatives are
\begin{equation}
    \frac{\partial\Omega_N}{\partial M_i^*} = \rho_s^i, \qquad \frac{\partial\Omega_N}{\partial\mu_i^*} = -\rho_v^i,
    \label{eq:dOmega}
\end{equation}
so that the $\sigma$ and $\delta$ equations acquire the sources $\sum_{i=p,n}(\partial M_i^*/\partial\varphi)\rho_s^i$, and the $\omega$ and $\rho$ equations acquire $g_\omega^N\sum_{i=p,n}\rho_v^i$ and $g_\rho^N\sum_{i=p,n}I_3^i\rho_v^i$. The equations of motion for the six mean fields then read
\begin{align}
    \frac{\partial\Omega}{\partial\sigma}:\quad &k_0\chi^2\sigma - 4k_1\left(\sigma^2+\zeta^2+\delta^2\right)\sigma - 2k_2\left(\sigma^3+3\sigma\delta^2\right) - 2k_3\chi\sigma\zeta - \frac{2\xi}{3}\frac{\chi^4\sigma}{\sigma^2-\delta^2} \nonumber\\
    & + \frac{\chi^2}{\chi_0^2}m_\pi^2 f_\pi + \sum_{i=p,n}\frac{\partial M_i^*}{\partial\sigma}\rho_s^i = 0, \label{eq:sigma}\\
    \frac{\partial\Omega}{\partial\zeta}:\quad &k_0\chi^2\zeta - 4k_1\left(\sigma^2+\zeta^2+\delta^2\right)\zeta - 4k_2\zeta^3 - k_3\chi\left(\sigma^2-\delta^2\right) - \frac{\xi}{3}\frac{\chi^4}{\zeta} \nonumber\\
    &+ \frac{\chi^2}{\chi_0^2}\left(\sqrt{2}\, m_K^2 f_K - \frac{m_\pi^2 f_\pi}{\sqrt{2}}\right) = 0, \label{eq:zeta}\\
    \frac{\partial\Omega}{\partial\delta}:\quad &k_0\chi^2\delta - 4k_1\left(\sigma^2+\zeta^2+\delta^2\right)\delta - 2k_2\left(\delta^3+3\sigma^2\delta\right) + 2k_3\chi\delta\zeta + \frac{2\xi}{3}\frac{\chi^4\delta}{\sigma^2-\delta^2} \nonumber\\
    & + \sum_{i=p,n}\frac{\partial M_i^*}{\partial\delta}\rho_s^i = 0, \label{eq:delta}\\
    \frac{\partial\Omega}{\partial\chi}:\quad &k_0\chi\left(\sigma^2+\zeta^2+\delta^2\right) - k_3\left(\sigma^2-\delta^2\right)\zeta + \left[1 + 4k_4 + 4\ln\frac{\chi}{\chi_0} - \frac{4\xi}{3}\ln\frac{\left(\sigma^2-\delta^2\right)\zeta}{\sigma_0^2\zeta_0}\right]\chi^3 \nonumber\\
    &+ \frac{2\chi}{\chi_0^2}\left[m_\pi^2 f_\pi\sigma + \left(\sqrt{2}\, m_K^2 f_K - \frac{m_\pi^2 f_\pi}{\sqrt{2}}\right)\zeta\right] - \frac{\chi}{\chi_0^2}\left(m_\omega^2\omega^2 + m_\rho^2\rho^2\right) = 0, \label{eq:chi}\\
    \frac{\partial\Omega}{\partial\omega}:\quad &-\frac{\chi^2}{\chi_0^2}m_\omega^2\omega - 4g_4\omega^3 - 12g_4\omega\rho^2 + g_\omega^N\sum_{i=p,n}\rho_v^i = 0, \label{eq:omega}\\
    \frac{\partial\Omega}{\partial\rho}:\quad &-\frac{\chi^2}{\chi_0^2}m_\rho^2\rho - 4g_4\rho^3 - 12g_4\omega^2\rho + g_\rho^N\sum_{i=p,n}I_3^i\rho_v^i = 0. \label{eq:rhoeq}
\end{align}
The $\zeta$ and $\chi$ equations carry no nucleon source, because the $u$ and $d$ quark masses of Eq. (\ref{eq:mud}), and hence $M_i^*$, contain neither $\zeta$ nor $\chi$. At given $\rho_B$, $T$, and $\eta$, we solve Eqs. (\ref{eq:sigma}) to (\ref{eq:rhoeq}) self-consistently together with Eq. (\ref{eq:eigen}) for the $u$ and $d$ quarks.

The nucleon source terms in Eqs. (\ref{eq:sigma}) and (\ref{eq:delta}) contain the derivatives of the effective nucleon masses with respect to the $\sigma$ and $\delta$ fields. We define the couplings of the nucleon to these two fields as
\begin{equation}
    g_\sigma^i = -\frac{\partial M_i^*}{\partial\sigma}, \qquad g_\delta^i = -\frac{\partial M_i^*}{\partial\delta},
    \label{eq:gN}
\end{equation}
so that the source terms read $-\sum_{i=p,n} g_\sigma^i \rho_s^i$ and $-\sum_{i=p,n} g_\delta^i \rho_s^i$, respectively. We evaluate both derivatives exactly through Eq. (\ref{eq:MN}), separately for the proton and the neutron. They are recomputed together with the mean fields at every $\rho_B$, $T$, and $\eta$, so that $g_\sigma^i$ and $g_\delta^i$ are density-dependent couplings. In the quark mean field model of Ref. \cite{Shen:2000}, which uses the confining potential of Eq. (\ref{eq:chic}), the in-medium $\sigma$-nucleon coupling is identified with the derivative of the effective nucleon mass with respect to $\sigma$. We extend the derivative prescription of Ref. \cite{Shen:2000} to the scalar-isovector field. We are not aware of an earlier calculation that obtains the nucleon-$\delta$ coupling from $-\partial M_i^*/\partial\delta$.

The chain rule through $m_u^*$ and $m_d^*$, worked out in Appendix \ref{app:gN}, gives the closed forms
\begin{equation}
    g_\sigma^i = \frac{g_s}{\sqrt{2}\, M_i^*}\sum_{q=u,d} n_q^i c_q^i, \qquad g_\delta^i = \frac{g_s}{\sqrt{2}\, M_i^*}\sum_{q=u,d} n_q^i I_3^q c_q^i, \qquad c_q^i = E_i r_q - \frac{1}{2}s_q,
    \label{eq:gclosed}
\end{equation}
where $r_q = de_q/dm_q^*$ and $s_q = d\langle p_q^2\rangle/dm_q^*$ are the derivatives of the quark energy and of its mean square momentum along Eq. (\ref{eq:eigen}), given in Eqs. (\ref{eq:dedm}) and (\ref{eq:dp2dm}). Here $n_u^p = n_d^n = 2$ and $n_d^p = n_u^n = 1$. The coupling $g_\delta^p$ is positive and $g_\delta^n$ is negative at every $\rho_B$, $T$, and $\eta$ considered here. Both couplings depend on $\rho_B$, $T$, and $\eta$ only through $\sigma$ and $\delta$, which fix $m_q^*$, $e_q$, and $M_i^*$. In isospin-asymmetric matter the $\delta$ field is finite, so that $m_u^* \neq m_d^*$ and, in general, $g_\sigma^p \neq g_\sigma^n$.
At $\eta = 0$, $g_\sigma^N$ decreases from $8.477$ in vacuum to $7.960$ at $\rho_B = \rho_0$ and $T = 0$, and equals $8.037$ at $\rho_B = \rho_0$ and $T = 0.1$ GeV. In pure neutron matter ($\eta = 0.5$) at $\rho_B = \rho_0$ and $T = 0$, we obtain $g_\sigma^p = 7.977$ and $g_\sigma^n = 7.968$, and $g_\delta^p = 1.347$ and $g_\delta^n = -1.311$.

The parameters of the model are listed in Table \ref{tab:param}. The decay constants $f_\pi$ and $f_K$ take the values of Ref. \cite{Wang:2004}. The pion and kaon masses enter Eq. (\ref{eq:LSB}) directly and are therefore reproduced by construction. For the dilaton vacuum value $\chi_0 = 282$ MeV, the coefficients $k_0$ to $k_4$ are fixed by five vacuum conditions. The $\sigma$, $\zeta$, and $\chi$ equations must hold at $\sigma_0$, $\zeta_0$, and $\chi_0$, the scalar meson mass must equal the value $m_\sigma = 417.5$ MeV of Ref. \cite{Wang:2001CTP}, and the sum of the squared $\eta$ and $\eta'$ masses must equal its physical value, $m_\eta^2 + m_{\eta'}^2 = (547^2 + 957^2)$ MeV$^2$, as in Refs. \cite{Lenaghan:2000ey,Sakai:2013nba}.
The coupling $g_s = 4.76$ gives the vacuum constituent quark mass $m_u^* = m_d^* = 313$ MeV, as in Ref. \cite{Wang:2004}, and $\Delta m_s$ is chosen to give $m_s^* = 450$ MeV in vacuum. The confinement strength $k_c$ and the spin correlation energies are fixed as described after Eqs. (\ref{eq:eigen}) and (\ref{eq:p2}). We then fit the nucleon-$\omega$ coupling $g_\omega^N$ and the vector coupling $g_4$ to the saturation point of symmetric nuclear matter at $T = 0$. The pressure $P = -\Omega$ must vanish there, and the energy per nucleon, measured from the vacuum nucleon mass, $E/A = (\Omega + \sum_{i=p,n}\mu_i\rho_v^i)/\rho_B - M_N$, must equal $-16.00$ MeV at $\rho_0 = 0.15$ fm$^{-3}$. The resulting parameter set, obtained with the density-dependent coupling $g_\sigma^i$ of Eq. (\ref{eq:gN}) in the $\sigma$ equation, differs from those of Refs. \cite{Singh:2024lra,Kaur:2025epja}. Finally, we fit the nucleon-$\rho$ coupling $g_\rho^N$ to the symmetry energy, $J = 32$ MeV at saturation.

\begin{table}[ht]
    \centering
    \renewcommand{\arraystretch}{1.3}
    \begin{tabular}{|c|c|c|c|}
        \hline
        Parameter & Value & Term & Constraint \\
        \hline
        $k_0$ & 4.5701 & \multirow{7}{*}{$\mathcal{L}_{\Sigma\Sigma}$} & $\partial\mathcal{L}_M/\partial\sigma|_{\rm vac} = 0$ \\
        $k_1$ & 2.6635 & & $\partial\mathcal{L}_M/\partial\zeta|_{\rm vac} = 0$ \\
        $k_2$ & $-9.9879$ & & $\partial\mathcal{L}_M/\partial\chi|_{\rm vac} = 0$ \\
        $k_3$ & $-4.3201$ & & $m_\sigma = 417.5$ MeV \cite{Wang:2001CTP} \\
        $k_4$ & $-0.1613$ & & $m_\eta^2 + m_{\eta'}^2 = (547^2 + 957^2)$ MeV$^2$ \\
        $\xi$ & 0.182 & & one-loop $\beta_{\rm QCD}$, $\xi \simeq 6/33$ \cite{Papazoglou:1998vr} \\
        $\chi_0$ & 282 MeV & & $K_0 \simeq 250$ MeV \\
        \hline
        $f_\pi$, $f_K$ & 93, 115 MeV & \multirow{2}{*}{$\mathcal{L}_{\chi SB}$} & Ref. \cite{Wang:2004} \\
        $m_\pi$, $m_K$ & 139, 494 MeV & & $\pi^\pm$, $K^\pm$ masses \\
        \hline
        $g_s$ & 4.76 & $\mathcal{L}_{qm}$ & $m_u^* = m_d^* = 313$ MeV in vacuum \\
        $\Delta m_s$ & $-11.11$ MeV & $\mathcal{L}_{\Delta m}$ & $m_s^* = 450$ MeV in vacuum \\
        $k_c$ & 300 MeV fm$^{-2}$ & $\mathcal{L}_c$ & $E/A$ and $P = 0$ at $\rho_0$ attainable \\
        \hline
        $E_{\rm spin}^i$ & $-567.012$ MeV & \multirow{2}{*}{Eq. (\ref{eq:EN})} & $M_N = 939.02$ MeV in vacuum \\
        $E_{\rm spin}^\Delta$ & $-309.956$ MeV & & $M_{\Delta^+} = 1232$ MeV in vacuum \\
        \hline
        $m_\omega$, $m_\rho$ & 783 MeV & $\mathcal{L}_{VV}$ & $\omega$ mass, $m_\rho = m_\omega$ \\
        $g_\omega^N$, $g_4$ & 11.8165, 65.5791 & Eq. (\ref{eq:mustar}), $\mathcal{L}_{VV}$ & $E/A = -16.00$ MeV, $P = 0$ at $\rho_0$ \\
        $g_\rho^N$ & 9.9922 & Eq. (\ref{eq:mustar}) & $J = 32$ MeV at $\rho_0$ \\
        \hline
    \end{tabular}
    \caption{Parameters of the CQMF model, the term of the Lagrangian or the equation in which each enters, and the constraint that fixes it. Vacuum denotes $\rho_B = T = 0$. The coefficients $k_0$ to $k_4$ are fixed together by the five conditions listed against them.}
    \label{tab:param}
\end{table}

The saturation properties of the model are obtained at $T = 0$ with the definitions of Ref. \cite{Kumar:2024MUSES}. In symmetric nuclear matter, $E/A$ has its minimum, with vanishing pressure, at the saturation density $\rho_0$, and its curvature there defines the incompressibility
\begin{equation}
    K_0 = 9\rho_0^2 \left.\frac{d^2(E/A)}{d\rho_B^2}\right|_{\rho_B = \rho_0}.
    \label{eq:K0}
\end{equation}
The symmetry energy $E_{\rm sym}(\rho_B)$ is taken as the difference between $E/A$ in pure neutron matter and in symmetric nuclear matter, and $J$ and $L$ are its value and slope at saturation,
\begin{equation}
    E_{\rm sym}(\rho_B) = \left.\frac{E}{A}\right|_{\eta = 0.5} - \left.\frac{E}{A}\right|_{\eta = 0}, \qquad J = E_{\rm sym}(\rho_0), \qquad L = 3\rho_0 \left.\frac{dE_{\rm sym}}{d\rho_B}\right|_{\rho_B = \rho_0}.
    \label{eq:JL}
\end{equation}
The values of these quantities are listed in Table \ref{tab:sat}, together with the ratio $M_N^*/M_N$ of the effective to the vacuum nucleon mass at $\rho_0$.
\begin{table}[ht]
    \centering
    \renewcommand{\arraystretch}{1.3}
    \begin{tabular}{|c|c|c|c|c|c|}
        \hline
        $\rho_0$ (fm$^{-3}$) & $E/A$ (MeV) & $K_0$ (MeV) & $J$ (MeV) & $L$ (MeV) & $M_N^*/M_N$ \\
        \hline
        0.150 & $-16.00$ & 250 & 32.01 & 67.67 & 0.684 \\
        \hline
    \end{tabular}
    \caption{Saturation properties of symmetric nuclear matter at $T = 0$ obtained with the parameters of Table \ref{tab:param}.}
    \label{tab:sat}
\end{table}

The medium enters the light-front calculation of Sec. \ref{sec:pdf} only through the effective masses obtained in this section. These are the constituent quark masses $m_u^*$ and $m_d^*$, the proton mass $M_p^*$, and the average mass $\langle M_\Delta^*\rangle$ of the four $\Delta$ charge states, each evaluated at the $\rho_B$, $T$, and $\eta$ of interest, and the diquark masses are built from them,
\begin{equation}
    M_S^* = m_u^* + m_d^*, \qquad
    M_V^* = \frac{3}{4}\,\frac{M_p^* + \langle M_\Delta^*\rangle}{2},
    \label{eq:MV}
\end{equation}
where a single axial-vector diquark mass is used for both the $ud$ and the $uu$ axial-vector diquarks. The factor $\frac{3}{4}$ follows Ref. \cite{Close:1988br}, where the diquark mass is three-quarters of the degenerate nucleon and $\Delta$ mass in the absence of one-gluon exchange. Here, the degenerate mass is replaced by the average of $M_p^*$ and $\langle M_\Delta^*\rangle$, and the one-gluon-exchange shifts of the diquark masses are not included. At zero density and temperature, Eq. (\ref{eq:MV}) gives $M_S = 0.626$ GeV and $M_V = 0.814$ GeV, the vacuum values used in Sec. III. The diquark masses carry no explicit $\delta$ term, so the $\delta$ field reaches them only through the splitting of $m_u^*$ and $m_d^*$ and the resulting shifts in $M_p^*$ and $\langle M_\Delta^*\rangle$. The proton and quark masses $M_p$ and $m_q$ of Sec. \ref{sec:pdf} take the in-medium values $M_p^*$ and $m_q^*$ obtained here.

\begin{figure*}[ht]
    \centering
    \begin{minipage}[c]{0.98\textwidth}
        (a)\includegraphics[width=4.7cm]{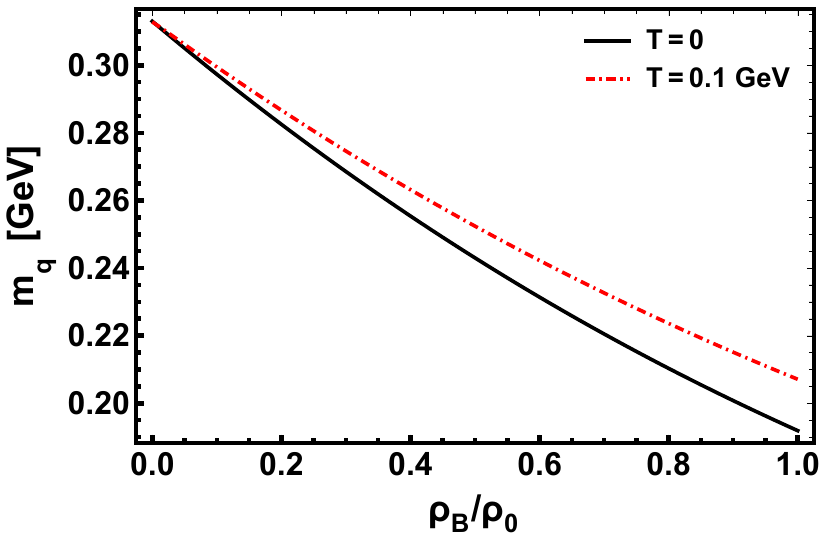}
        \hspace{0.03cm}
        (b)\includegraphics[width=4.7cm]{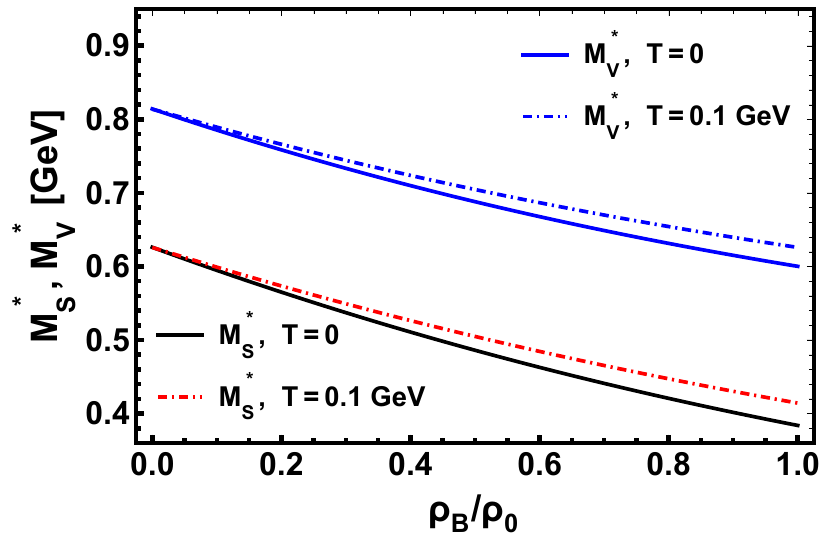}
        \hspace{0.03cm}
        (c)\includegraphics[width=4.7cm]{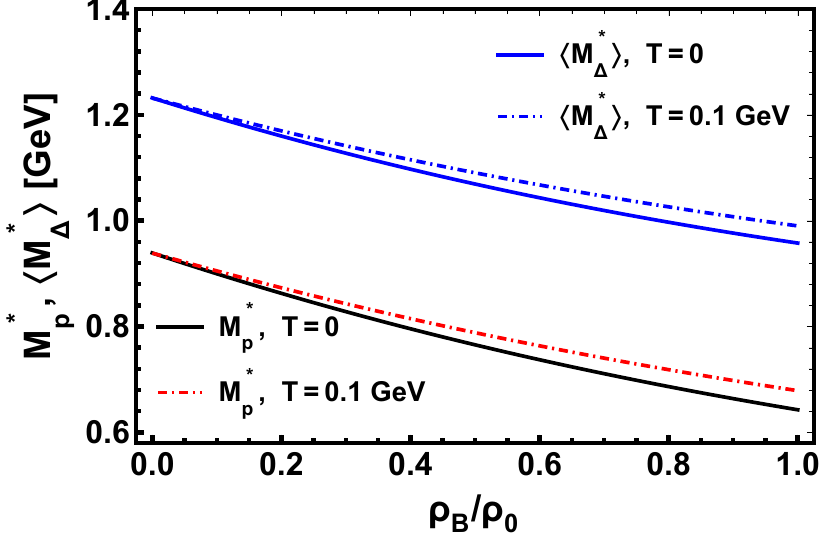}
    \end{minipage}
    \caption{Effective masses of (a) the constituent quarks, $m_q^* = m_u^* = m_d^*$, (b) the scalar and axial-vector diquarks, $M_S^* = m_u^* + m_d^*$ and $M_V^*$ of Eq. (\ref{eq:MV}), and (c) the proton, $M_p^*$, and the average $\Delta$ baryon mass $\langle M_\Delta^*\rangle$, as functions of the baryon density $\rho_B/\rho_0$ in symmetric nuclear matter ($\eta = 0$) at $T = 0$ (solid lines) and $T = 0.1$ GeV (dash-dotted lines). The shifts of these masses between $\eta = 0$ and $\eta = 0.5$ are listed in Table \ref{tab:dmass}.}
    \label{fig:mstar}
\end{figure*}

The effective masses of the constituent quarks, the scalar and axial-vector diquarks, the proton, and the average $\Delta$ baryon in symmetric nuclear matter are plotted as functions of the baryon density in Fig. \ref{fig:mstar}. We begin with $T = 0$, where all five masses decrease monotonically as $\rho_B$ increases from zero to $\rho_0$, but by different fractions. The constituent quark mass decreases from $m_q^* = 0.313$ GeV in vacuum to $0.192$ GeV at $\rho_0$, a drop of about $39\%$. The scalar diquark mass $M_S^* = m_u^* + m_d^*$ is twice the quark mass at $\eta = 0$ and follows it from $0.626$ GeV to $0.384$ GeV. The proton mass decreases by about $32\%$, from $0.939$ GeV to $0.643$ GeV, so that $M_N^*/M_N = 0.684$, as listed in Table \ref{tab:sat}. The axial-vector diquark mass decreases by about $26\%$, from $0.814$ GeV to $0.600$ GeV, less than the proton mass. This can be traced to Eq. (\ref{eq:MV}), where $M_V^*$ is three-quarters of the mean of $M_p^*$ and $\langle M_\Delta^*\rangle$, and $\langle M_\Delta^*\rangle$ falls by about $22\%$, from $1.232$ GeV to $0.958$ GeV, the smallest fraction of the five masses.

At $T = 0.1$ GeV (dashed lines in Fig. \ref{fig:mstar}), the masses again decrease monotonically with $\rho_B$ but lie above their $T = 0$ values. For example, at $\rho_B = \rho_0$ the constituent quark mass is $m_q^* = 0.207$ GeV at $T = 0.1$ GeV, against $0.192$ GeV at $T = 0$, which is higher by about $8\%$. The scalar diquark mass rises by the same fraction, to $0.414$ GeV, since it remains twice the quark mass in symmetric matter. The proton mass rises by about $6\%$, to $0.679$ GeV, and the axial-vector diquark mass by about $4\%$, to $0.626$ GeV. As at $T = 0$, the smaller change of $M_V^*$ can be traced to Eq. (\ref{eq:MV}), since $\langle M_\Delta^*\rangle$ rises by only about $3\%$, from $0.958$ GeV to $0.990$ GeV. The temperature, therefore, partly offsets the density-driven decrease of the effective masses. The masses that fall most with density, $m_q^*$ and $M_S^*$, also rise most with temperature, followed by $M_p^*$, $M_V^*$, and $\langle M_\Delta^*\rangle$.

In asymmetric matter, the $\delta$ field is sourced by $-\sum_{i=p,n} g_\delta^i\rho_s^i$ in Eq. (\ref{eq:delta}), and since $g_\delta^p > 0$ and $g_\delta^n < 0$, this source follows the imbalance between the proton and neutron scalar densities. It vanishes in symmetric nuclear matter and is finite at $\eta = 0.5$, which corresponds to pure neutron matter. Through Eq. (\ref{eq:mud}), $\delta$ shifts $m_u^*$ and $m_d^*$ by equal amounts in opposite directions and drops out of $m_u^* + m_d^*$, which depends on $\sigma$ alone. The resulting relative changes from $\eta = 0$ to $\eta = 0.5$ are listed in Table \ref{tab:dmass}. At $T = 0$, the $\sigma$ field also changes with $\eta$ and lifts both quark masses by $1.26\%$ at $\rho_0$, so $m_u^*$ rises by $2.69\%$ while $m_d^*$ falls by only $0.17\%$. At $T = 0.1$ GeV, the change of $\sigma$ stays within $0.11\%$, and the splitting is close to symmetric, with $1.25\%$ for $m_u^*$ and $-1.03\%$ for $m_d^*$ at $\rho_0$. The scalar diquark mass $M_S^* = m_u^* + m_d^*$ does not depend on $\delta$ and changes by at most $1.26\%$ at $T = 0$ and $0.11\%$ at $T = 0.1$ GeV, while $M_V^*$ and $M_p^*$ change by at most $0.81\%$ and $1.23\%$ at $T = 0$ and $0.17\%$ and $0.36\%$ at $T = 0.1$ GeV. The isospin asymmetry has the smallest effect on the masses, at most $2.69\%$ in Table \ref{tab:dmass}, for $m_u^*$ at $\rho_0$ and $T = 0$, compared with the decrease of $m_q^*$ by about $39\%$ from vacuum to $\rho_0$ and its increase by about $8\%$ at $\rho_0$ from $T = 0$ to $T = 0.1$ GeV.

\begin{table}[ht]
    \centering
    \renewcommand{\arraystretch}{1.3}
    \setlength{\tabcolsep}{10pt}
    \begin{tabular}{|c|c|c|c|c|c|c|}
        \hline
        \multirow{2}{*}{$T$ (GeV)} & \multirow{2}{*}{$\rho_B$} & \multicolumn{5}{c|}{$\Delta m^*/m^*$ (\%)} \\ \cline{3-7}
        & & $m_u^*$ & $m_d^*$ & $M_S^*$ & $M_V^*$ & $M_p^*$ \\ \hline
        \multirow{4}{*}{0} & 0.25$\rho_0$ & 0.34 & $-0.17$ & 0.08 & 0.09 & 0.14 \\ \cline{2-7}
        & 0.5$\rho_0$ & 0.91 & $-0.26$ & 0.32 & 0.26 & 0.40 \\ \cline{2-7}
        & 0.75$\rho_0$ & 1.70 & $-0.27$ & 0.72 & 0.52 & 0.78 \\ \cline{2-7}
        & $\rho_0$ & 2.69 & $-0.17$ & 1.26 & 0.81 & 1.23 \\ \hline
        \multirow{4}{*}{0.1} & 0.25$\rho_0$ & 0.17 & $-0.25$ & $-0.04$ & 0.00 & 0.03 \\ \cline{2-7}
        & 0.5$\rho_0$ & 0.46 & $-0.50$ & $-0.02$ & 0.04 & 0.11 \\ \cline{2-7}
        & 0.75$\rho_0$ & 0.82 & $-0.77$ & 0.03 & 0.10 & 0.22 \\ \cline{2-7}
        & $\rho_0$ & 1.25 & $-1.03$ & 0.11 & 0.17 & 0.36 \\ \hline
    \end{tabular}
    \caption{Relative change of the effective quark, diquark and proton masses from symmetric ($\eta = 0$) to asymmetric ($\eta = 0.5$) nuclear matter, $\Delta m^*/m^* = [m^*(\eta = 0.5) - m^*(\eta = 0)]/m^*(\eta = 0)$, at the listed temperatures and baryon densities.}
    \label{tab:dmass}
\end{table}

\section{Parton distribution functions \label{sec:pdf}}
In this section, we investigate the implications of medium effects on the valence quark structure of the proton through the PDFs. We  choose a frame where the proton carries zero transverse momentum, and its four-momentum in vacuum condition is given by 
\begin{eqnarray}
    P=\bigg[P^+, \frac{M^2_p}{2P^+}, \textbf{0}_T \bigg],
\end{eqnarray}
using the generic light-front four-vector notation $v=[v^+,v^-,\textbf{v}_T]$, where $M_p$ denotes the proton mass and the four-momentum of the active quark $q$ is defined as
\begin{eqnarray}
    k=\bigg[xP^+, \frac{k^2+\textbf{k}^2_T}{2xP^+}, \textbf{k}_T \bigg].
\end{eqnarray}
Here, the longitudinal momentum fraction carried by an active quark is expressed by $x=k^+/P^+$. For semi-inclusive deep inelastic scattering, the integrated quark-quark correlator is defined as
\begin{equation}
    \Phi (x,k_T;S) = \int \frac{d \xi^- d^2 \xi_T}{(2\pi)^3} e^{ik\cdot \xi} \bra{P;S} \bar{\psi}(0) W_{[0,\xi]} \psi(\xi) \ket{P;S} \bigg|_{\xi^+=0},
\end{equation} 
with $\Gamma=\{1,\gamma^\mu,\sigma_{\mu \nu}, i\gamma_5, \gamma_5\gamma^\mu\}$ \cite{Meissner:2009ww}. Wilson line $W$ preserves the gauge invariance that connects the points 0 and $\xi$ with an infinitely long staple-shaped path along the lightlike direction $n$. In a Lorentz covariant fashion, the light-front decomposition of a vector $n$ can be written in two light-like vectors $n_+ = [1,0,\textbf{0}_T]$ and $n_- = [0,1,\textbf{0}_T]$. In the light-front gauge, $A^+=0$, the Wilson line reduces to unity. At the leading twist, the integration over $k_T$ leads to the collinear quark correlator, whose leading-twist parametrization involves the familiar collinear distributions $f_1^q(x)$, $g_1^q(x)$, and $h_1^q(x)$ \cite{Bacchetta:2006tn}, which is defined as
\begin{equation}
    \Phi(x;S)=\frac{1}{2} \big[f_1^q(x) \slashed{n}_+ + g_1^q(x) S_L \gamma_5 \slashed{n}_+ + h_1^q(x) i\sigma_{\mu \nu} \gamma_5 n^\mu_+ S^\nu_T \big].
\end{equation}
The spin vector $S^\mu$ of the proton target can be defined in terms of  the longitudinal and transverse components $\{S_L, S_T\}$ as
\begin{equation}
    S^\mu = S_L \frac{(P \cdot n_-) n_+^\mu - (P \cdot n_+) n_-^\mu}{M_p}+S_T^\mu.
\end{equation}
We evaluate this $k_T$ integrated quark-quark correlator in the spectator approximation, where we consider a proton as a two-body system of an active quark $q$ and a spectator diquark $D$. Depending on the spin of spectator diquark $S_D=\{0,1\}$, it is termed as scalar $S$ and axial-vector $V$ diquarks, respectively. Further, isospin categorizes the axial-vector diquarks as isoscalar ($ud$-like system) and isovector ($uu$-like system) \cite{Jakob:1997wg}. So, assuming spin-isospin symmetry, the probabilistic weights among scalar-isoscalar, axial-vector-isoscalar, and axial-vector-isovector are 3:1:2 as the proton wave function \cite{Lichtenberg:1968zz} is written as 
\begin{eqnarray}
    \ket{p}^\lambda=\frac{1}{\sqrt 2} \ket{u S(ud)}^\lambda - \frac{1}{\sqrt 6} \ket{u V(ud)}^\lambda + \frac{1}{\sqrt 3} \ket{d V(uu)}^\lambda.
\end{eqnarray}
This choice leads to the relations $f^u=3f^{u(S)}/2 + f^{u(V)}/2$ and $f^d=f^{d(V)}$, where $f=\{f_1(x), g_1(x), h_1(x)\}$. On parameterization, the PDFs can be written in terms of the overlap form of respective scalar $\psi^\lambda_{\lambda_q}$ and axial-vector $\psi^\lambda_{\lambda_q \lambda_D}$ LFWFs, where $\lambda_q=\{\uparrow,\downarrow\}$ and $\lambda_D=\{\pm1,0\}$ denote the helicities of an active quark and a spectator diquark, sequentially. The contribution of the scalar diquark is given by
\begin{eqnarray}
    f_1^{q(S)} (x) &=& \int \frac{d^2 \textbf{k}_T}{16 \pi^3} \big[|\psi^\Uparrow_{\uparrow}(x,\textbf{k}_T)|^2 + |\psi^\Uparrow_{\downarrow} (x,\textbf{k}_T)|^2 \big], \\
    g_1^{q(S)} (x) &=& \int \frac{d^2 \textbf{k}_T}{16 \pi^3} \big[|\psi^\Uparrow_{\uparrow} (x,\textbf{k}_T)|^2 - |\psi^\Uparrow_{\downarrow} (x,\textbf{k}_T)|^2 \big], \\
    h_1^{q(S)} (x) &=& \int \frac{d^2 \textbf{k}_T}{16 \pi^3} \big[\psi^{\Uparrow \ast}_{\uparrow} (x,\textbf{k}_T) \psi^\Downarrow_{\downarrow} (x,\textbf{k}_T) - \psi^{\Uparrow \ast}_{\downarrow}(x,\textbf{k}_T) \psi^\Downarrow_{\uparrow} (x,\textbf{k}_T) \big]. 
\end{eqnarray}
On the parallel lines, the contribution of the axial-vector diquark is given by
\begin{eqnarray}
    f_1^{q(V)} (x) &=& \int \frac{d^2 \textbf{k}_T}{16 \pi^3} \big[|\psi^\Uparrow_{\uparrow 1}(x,\textbf{k}_T)|^2 + |\psi^\Uparrow_{\uparrow 0}(x,\textbf{k}_T)|^2 + |\psi^\Uparrow_{\uparrow -1}(x,\textbf{k}_T)|^2 + |\psi^\Uparrow_{\downarrow 1} (x,\textbf{k}_T)|^2 \nonumber \\ &+& |\psi^\Uparrow_{\downarrow 0} (x,\textbf{k}_T)|^2+ |\psi^\Uparrow_{\downarrow -1} (x,\textbf{k}_T)|^2\big], \\
    g_1^{q(V)} (x) &=& \int \frac{d^2 \textbf{k}_T}{16 \pi^3} \big[|\psi^\Uparrow_{\uparrow 1}(x,\textbf{k}_T)|^2 + |\psi^\Uparrow_{\uparrow 0}(x,\textbf{k}_T)|^2 + |\psi^\Uparrow_{\uparrow -1}(x,\textbf{k}_T)|^2 -( |\psi^\Uparrow_{\downarrow 1} (x,\textbf{k}_T)|^2 \nonumber \\ &+& |\psi^\Uparrow_{\downarrow 0} (x,\textbf{k}_T)|^2+ |\psi^\Uparrow_{\downarrow -1} (x,\textbf{k}_T)|^2) \big], \\
    h_1^{q(V)} (x) &=& \int \frac{d^2 \textbf{k}_T}{16 \pi^3} \big[\psi^{\Uparrow \ast}_{\uparrow 1} (x,\textbf{k}_T) \psi^\Downarrow_{\downarrow 1} (x,\textbf{k}_T) + \psi^{\Uparrow \ast}_{\uparrow 0} (x,\textbf{k}_T) \psi^\Downarrow_{\downarrow 0} (x,\textbf{k}_T) 
    + \psi^{\Uparrow \ast}_{\uparrow -1} (x,\textbf{k}_T) \psi^\Downarrow_{\downarrow -1} (x,\textbf{k}_T) \nonumber \\ &+& \psi^{\Uparrow \ast}_{\downarrow 1}(x,\textbf{k}_T) \psi^\Downarrow_{\uparrow 1} (x,\textbf{k}_T) + \psi^{\Uparrow \ast}_{\downarrow 0}(x,\textbf{k}_T) \psi^\Downarrow_{\uparrow 0} (x,\textbf{k}_T) + \psi^{\Uparrow \ast}_{\downarrow -1}(x,\textbf{k}_T) \psi^\Downarrow_{\uparrow -1} (x,\textbf{k}_T) \big]. 
\end{eqnarray}
The required LFWFs are computed by using
\begin{eqnarray}
    \psi^{\lambda}_{\lambda_q} (x,\bfk) &=& \sqrt{\frac{k^+}{(P-k)^+}} \frac{1}{k^2-m^2_q} \bar{u} (k,\lambda_q) ~\mathcal{Y}_{S}~ U(P,\lambda), \\
    \psi^{\lambda}_{\lambda_q \lambda_V}  (x,\bfk) &=& \sqrt{\frac{k^+}{(P-k)^+}} \frac{1}{k^2-m^2_q} \bar{u} (k,\lambda_q) \epsilon^\ast_\mu (P-k,\lambda_V) \cdot \mathcal{Y}_V^\mu~ U(P,\lambda),
    \label{VectorWfn}
    \label{ScalarWfn}
\end{eqnarray}
with  scalar and axial-vector vertex $\mathcal{Y}_S=ig_{S}(k^2) \bf{1}$ and $\mathcal{Y}_V^\mu=ig_{V}(k^2) \gamma^\mu \gamma_5 /\sqrt{2}$, sequentially \cite{Bacchetta:2008af}. The dipolar form factor for spectator diquarks has a form 
\begin{eqnarray}
    g_{qD}(k^2) = g_{qD} \frac{k^2-m_q^2}{|k^2-\Lambda^2_{qD}|^2}.
\end{eqnarray}
Here, the $\Lambda_{qD}$ and $g_{qD}$ are the cutoff parameter and coupling constant, respectively. The momentum space wave function in the spectator diquark model is defined as
\begin{eqnarray}
    \phi_{qD} =- \frac{g_{qD}}{\sqrt{1-x}} \frac{x (1-x)^2}{(\bfk^2+L^2_{qD})^2} \, ,
    \label{MSWF}
\end{eqnarray}
with $L^2_{qD}=x M^2_D+(1-x) \Lambda_{qD}^2-x(1-x)M_p^2$, where $M_D$ is the mass of the spectator diquark, $M_S$ for the scalar and $M_V$ for the axial-vector diquark. Here, 
$\Lambda_{qD}$ is an appropriate cutoff to avoid the singularity and satisfy the relation $M_D>M_p-\Lambda_{qD}$. The explicit expressions of the LFWFs are expressed in Table \ref{LFWFs}. The vacuum numerical values of proton and quark mass are considered as $M_p=0.939$ and $m_q=0.313$ in the units of GeV. 
The diquark masses are those of Eq. (\ref{eq:MV}) in Section \ref{Sec2}, $M_S=0.626$ GeV and $M_V=0.814$ GeV in vacuum.
The cutoff parameters $\Lambda_{uS(ud)}=0.5$ GeV, $\Lambda_{uV(ud)}=0.6$ GeV, and $\Lambda_{dV(uu)}=0.4$ GeV are chosen in such a way that they give the value of axial-charge of proton as close as to its experimental value, i.e., $1.27$ \cite{Leader:2010rb, Markisch:2018ndu}. The original couplings $g_D$
are rescaled by normalization constants $N_{qD}$ \cite{Bacchetta:2008af} and the $f_1^q$ TMD is normalized by
\begin{eqnarray}
    \int dx \int d^2 \bfk ~ f^q_{1norm}(x,\bfk) = 1.
\end{eqnarray}
In the current section, the formalism and LFWFs discussed above are formulated under vacuum conditions at model scale $Q_0^2=0.2$ GeV$^2$. In order to incorporate medium effects, the corresponding vacuum masses of the quarks, diquarks, and proton are replaced by their baryon density, temperature, and isospin asymmetry dependent in-medium values, along with satisfying the normalization condition at each fixed value of $\{\rho_B, T, \eta\}$. The cutoffs $\Lambda_{qD}$ keep their vacuum values.
\begin{table}[h]
    \centering
    \begin{tabular}{|c|c|c|c|c|}
        \hline
        $~~~~~~D~~~~~~$ & $~~~~~\lambda_q~~~~~$ & $~~~~~\lambda_{_D}~~~~~$ & $\lambda=\Uparrow$  & $\lambda=\Downarrow$ \\
        \hline
        \multirow{2}{*}{$~~S~~$} & $~~+\frac{1}{2}~~$ & $-$ & $\frac{m_q+x M_p}{x} \, \phi_{{qS}}$ & $\frac{k_x-i k_y}{x}  \phi_{{qS}}$ \\
        & $~~-\frac{1}{2}~~$ & $-$ & $-\frac{k_x+i k_y}{x}  \phi_{{qS}}$ & $\frac{m_q+x M_p}{x} \, \phi_{{qS}}$ \\
        \hline
        \multirow{6}{*}{$~~V~~$} & $~~+\frac{1}{2}~~$ & $+1$ & $\frac{k_x-i k_y}{x(1-x)} \, \phi_{{qV}}$ & 0 \\
        & $~~-\frac{1}{2}~~$ & $+1$ & $\frac{m_q+x M_p}{x} \, \phi_{{qV}}$ & $- \frac{k_x-i k_y}{(1-x)}  \phi_{{qV}}$ \\
        & $~~+\frac{1}{2}~~$ & $0$ & $~~~~\frac{\bfk^2-x M_V^2-m_qM_p(1-x)^2}{\sqrt{2}x(1-x)M_V} \, \phi_{{qV}}~~~~$ & $\frac{(m_q+M_p)(k_x-i k_y)}{\sqrt{2}x M_V}  \phi_{{qV}}$  \\
        & $~~-\frac{1}{2}~~$ & $0$ & $\frac{(m_q+M_p)(k_x+i k_y)}{\sqrt{2}x M_V}  \phi_{{qV}}$ & $-\frac{\bfk^2-x M_V^2-m_qM_p(1-x)^2}{\sqrt{2}x(1-x) M_V} \, \phi_{{qV}}$ \\
        & $~~+\frac{1}{2}~~$ & $-1$ & $- \frac{k_x+i k_y}{(1-x)}  \phi_{{qV}}$ & $-\frac{m_q+x M_p}{x} \, \phi_{{qV}}$ \\
        & $~~-\frac{1}{2}~~$ & $-1$ & 0 & $\frac{k_x+i k_y}{x(1-x)} \, \phi_{{qV}}$ \\
        \hline
    \end{tabular}
    \caption{The light-front wave functions of Ref. \cite{Bacchetta:2008af} for a proton carrying helicity $\lambda$, comprised of an active quark and a spectator diquark with helicities $\lambda_q$ and $\lambda_D$, respectively, used in the present calculations with $\phi_{qD}$ as momentum space wave functions. $M_p$, $m_q$, and $M_D$ are the masses of the proton, active quark, and spectator diquark, which take their in-medium values from Sec. \ref{Sec2}.}
    \label{LFWFs}
\end{table}
\subsection{Baryon density effects on parton distribution functions}
\begin{figure*}
    \centering
    \begin{minipage}[c]{0.98\textwidth}
        (a)\includegraphics[width=4.7cm]{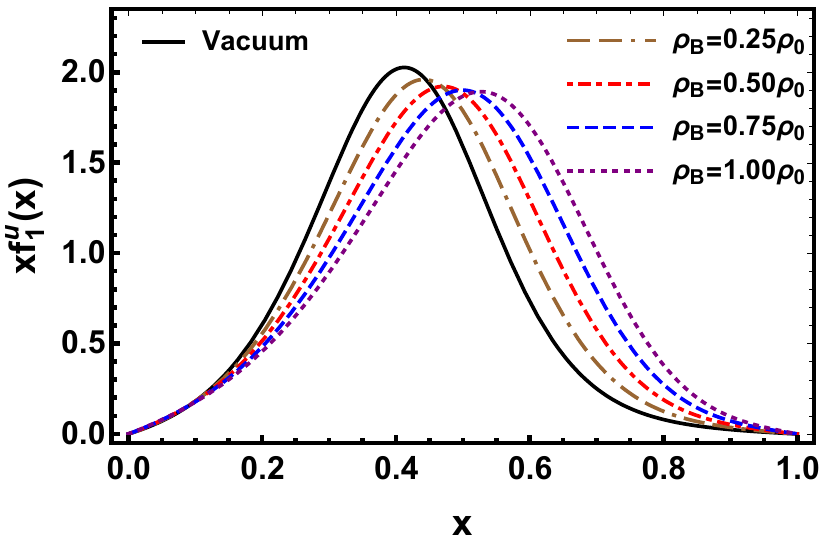}
        \hspace{0.03cm}
        (b)\includegraphics[width=4.7cm]{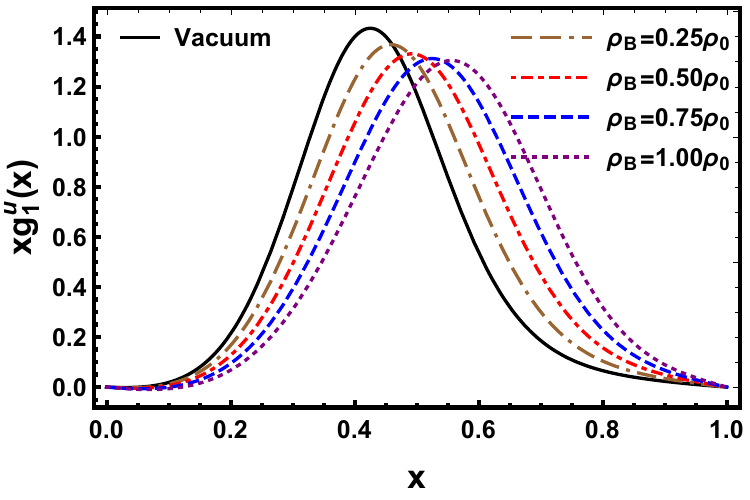}
        \hspace{0.03cm}	 
        (c)\includegraphics[width=4.7cm]{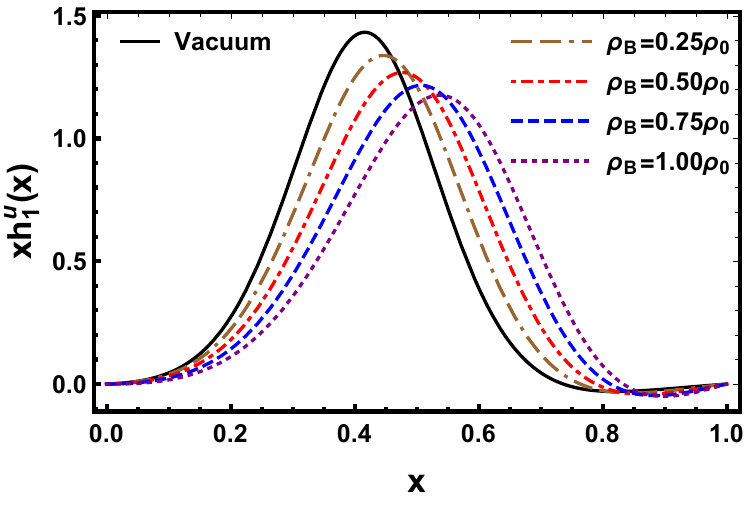}
        \hspace{0.03cm}
        (d)\includegraphics[width=4.7cm]{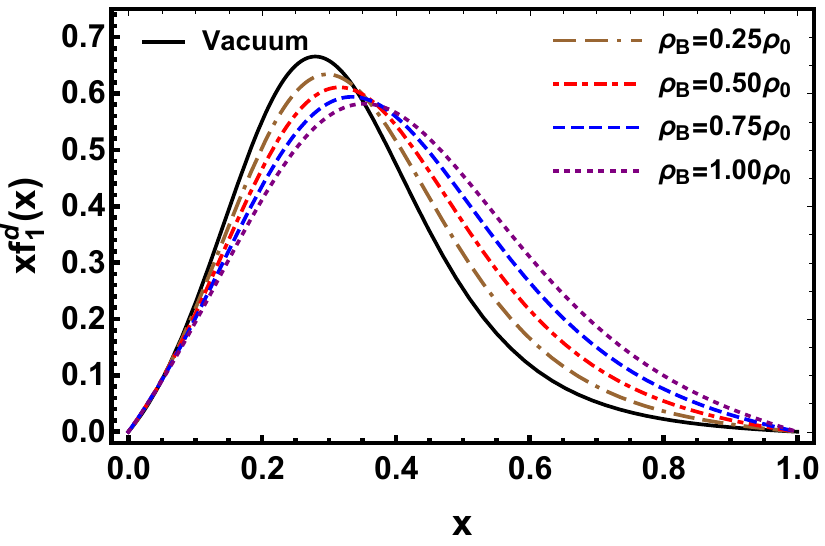}
        \hspace{0.03cm}	
        (e)\includegraphics[width=4.7cm]{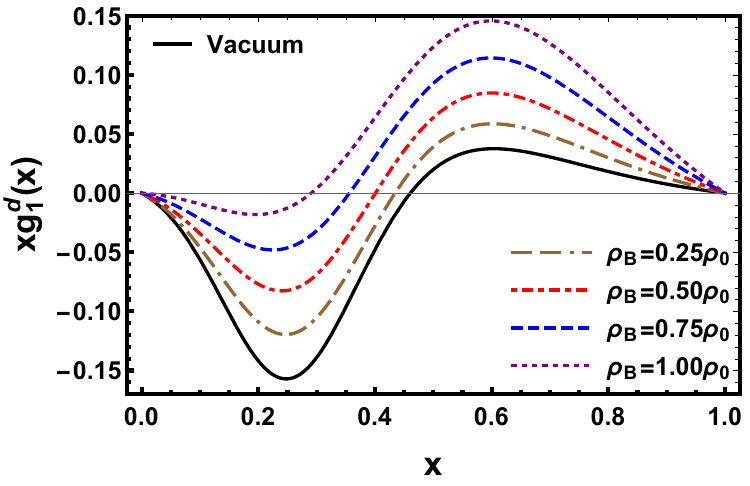}
        \hspace{0.03cm}
        (f)\includegraphics[width=4.7cm]{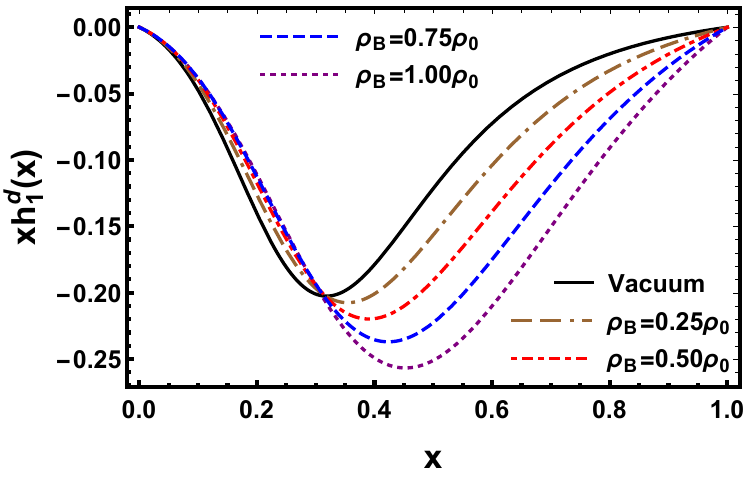}
        \hspace{0.03cm}
    \end{minipage}
    \caption{\label{figBD} The leading-twist proton parton distribution functions as a function of the longitudinal momentum fraction $x$ for different baryon densities $\rho_B$ at zero temperature and isospin asymmetry. The top (bottom) row corresponds to the $u$ ($d$) quark distributions, while the first, second, and third columns represent the unpolarized distribution $xf_1(x)$, helicity distribution $xg_1(x)$, and transversity distribution $xh_1(x)$, respectively.}
\end{figure*}

A set of PDFs is presented as a function of longitudinal momentum fraction $x$ for discrete values of baryon density $\rho_B$ in Fig. \ref{figBD}. The unpolarized PDFs for $f_1^u$ and $f_1^d$ are shown in Figs. \ref{figBD}(a) and \ref{figBD}(d), respectively. With increasing baryon density, the distributions exhibit a suppression of their peak amplitudes, accompanied by a broadening of the $x$-dependence and a shift of the peak toward larger $x$. This behavior reflects the medium-induced redistribution of the longitudinal momentum fraction of quarks, while the valence-quark number remains conserved through the normalization of the corresponding first moment at each baryon density. The peak amplitude of $u$($d$) quark decreases by $6.6\%$ ($12.6\%$), while the peak position shifts toward higher $x$ by $27.9\%$ ($26\%$). The second Mellin moment of unpolarized PDF is tabulated in the Table \ref{tabfq} for different values of baryon density. For the vacuum case, the sum of the second moment of $f_1^u$ and $f_1^d$ gives 0.926, which is not exactly unity because in our spectator model, it is not possible to fulfill at the same time the momentum sum rule and the quark number sum rule \cite{Bacchetta:2008af}. The same limitation persists in the in-medium extension of the model.
The results of the in-medium second Mellin moment show a gradual increase in the Mellin moment with increasing baryon density. This behavior can be attributed to the rapid decrease in the quark, diquark, and proton masses with increasing baryon density, as our calculation incorporates only the corresponding effective masses. A similar trend has been reported in Ref. \cite{Mineo:2003vc}, where the use of effective masses alone leads to a shift of the unpolarized PDF peak toward higher $x$ value. Our results are therefore consistent with the scalar-field only results of Ref. \cite{Mineo:2003vc}, where the distribution becomes stiffer. In that work, the quenching in the valence region is also reported to arise from the vector field, which is not included here.

The results of helicity distribution for $u$ and $d$ quarks are shown in the respective Figs. \ref{figBD}(b) and \ref{figBD}(e). The $u$ quark distribution shows a gradual suppression of its peak, from 1.43 in vacuum to 1.30 at saturation density, with a shift of the peak position from $x=0.42$ to $x=0.56$ (see Fig. \ref{figBD}(b)). In the vacuum, the $d$ quark distribution becomes positive as $x\rightarrow1$. This behavior is consistent with the increasing contribution of the nonzero orbital angular momentum components of the LCWFs at large $x$ \cite{Avakian:2007xa}. With increasing baryon density, the positive large-$x$ component of the $d$ quark helicity distribution becomes progressively enhanced, while its negative contribution at small $x$ is suppressed. Consequently, the zero-cross over point shifts toward smaller $x$. This behavior can be associated with the medium-induced modification of the LFWFs, where the density-dependent quark, diquark, and proton masses alter the relative contributions of the different orbital angular momentum components to the helicity distribution. First Mellin moment of the helicity distribution, i.e., axial charge, is tabulated in the Table \ref{tabAC}, corresponding to both quarks for discrete values of the baryon density. The magnitude of the axial charge for both $\Delta u$ and $\Delta d$ is found to decrease with an increase in the baryon density, which is consistent qualitatively with the result of Ref. \cite{Cloet:2006bq}. For the case of proton, the axial charge $g_A=\Delta u - \Delta d$ drops from 1.28 value in vacuum to 0.844 at saturation density. In particular, the reduction of the axial charge suggests that the quark helicity contribution to the proton spin becomes smaller in the medium. 

The results of transversity distribution for $u$ and $d$ quark as a function of longitudinal momentum fraction $x$ for discrete values of the baryon density $\rho_B$ are presented in Figs. \ref{figBD}(c) and \ref{figBD}(f), respectively. With increasing baryon density from $\rho_B=0$ to the saturation density, the qualitative behavior of both distributions remains unchanged.
The $u$ quark peak decreases while the $d$ quark peak deepens, and both move to higher $x$.
Quantitatively, the $u$ quark peak decreases from 1.43 at $x=0.42$ to 1.18 at $x=0.54$ when $\rho_B$ increases from zero to saturation density. For the $d$ quark distribution, the corresponding negative peak value changes from $-0.20$ at $x=0.32$ to $-0.26$ at $x=0.45$. The first Mellin moments, corresponding to the tensor charges, exhibit a decrease in the $u$ quark contribution, while the $d$ quark contribution becomes increasingly negative with increasing baryon density. The corresponding values of the tensor charge $\delta q$ at discrete values of the baryon density are listed in the Table \ref{tabTC}.
In a nutshell, the decrease in the tensor charge reflects a corresponding modification of the transverse spin structure.
\subsection{Temperature effects on parton distribution functions}
\begin{figure*}
    \centering
    \begin{minipage}[c]{0.98\textwidth}
        (a)\includegraphics[width=4.7cm]{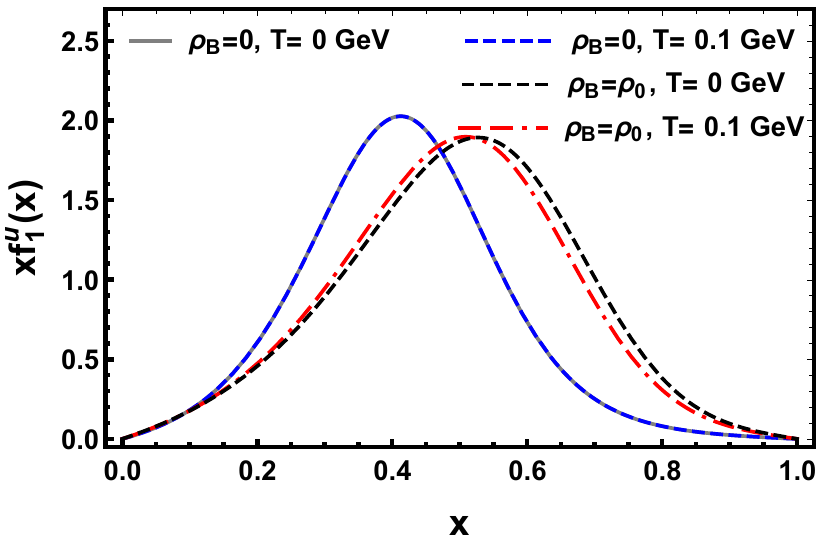}
        \hspace{0.03cm}
        (b)\includegraphics[width=4.7cm]{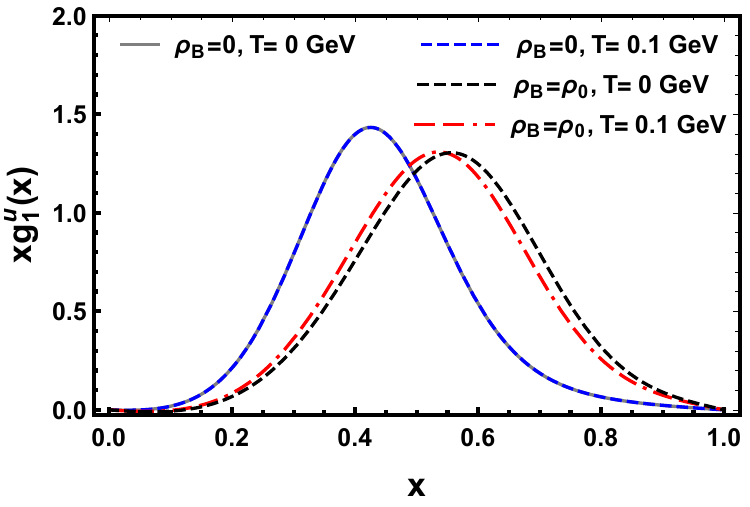}
        \hspace{0.03cm}
        (c)\includegraphics[width=4.7cm]{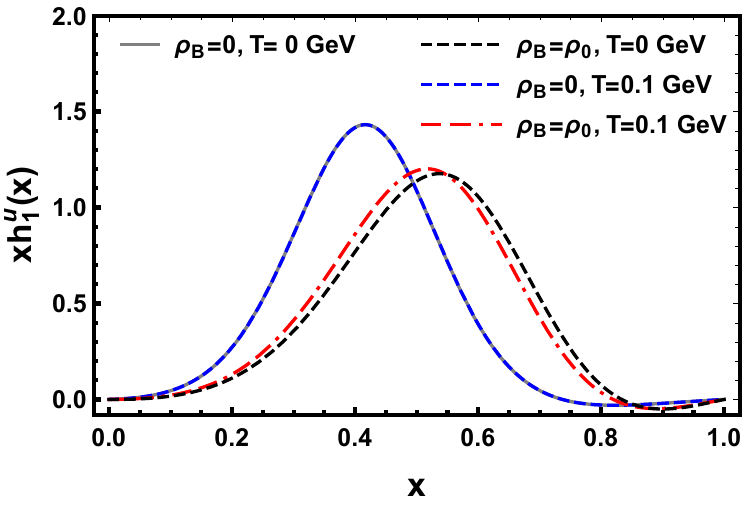}
        \hspace{0.03cm}
        (d)\includegraphics[width=4.7cm]{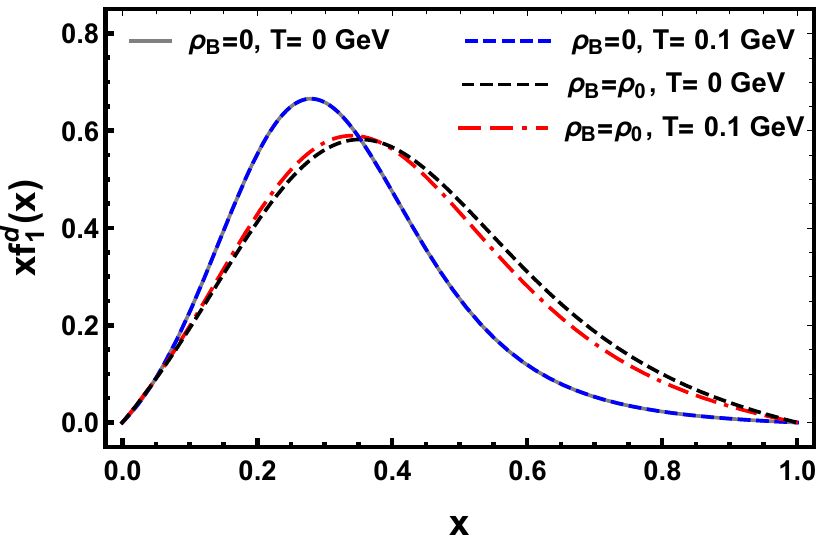}
        \hspace{0.03cm}	
        (e)\includegraphics[width=4.7cm]{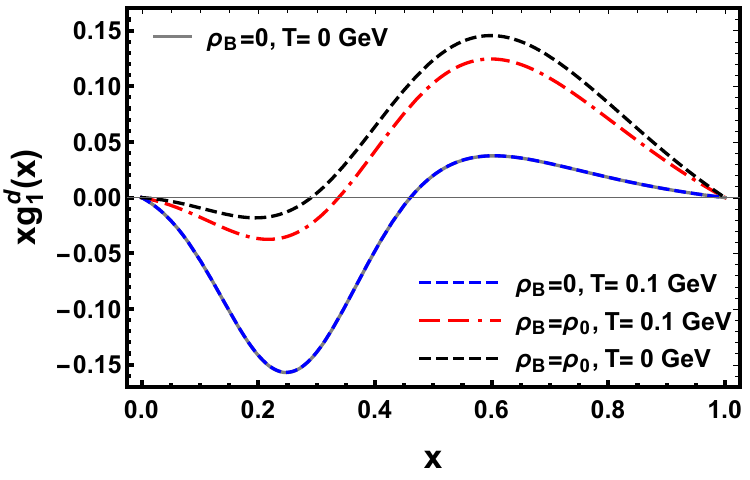}
        \hspace{0.03cm}
        (f)\includegraphics[width=4.7cm]{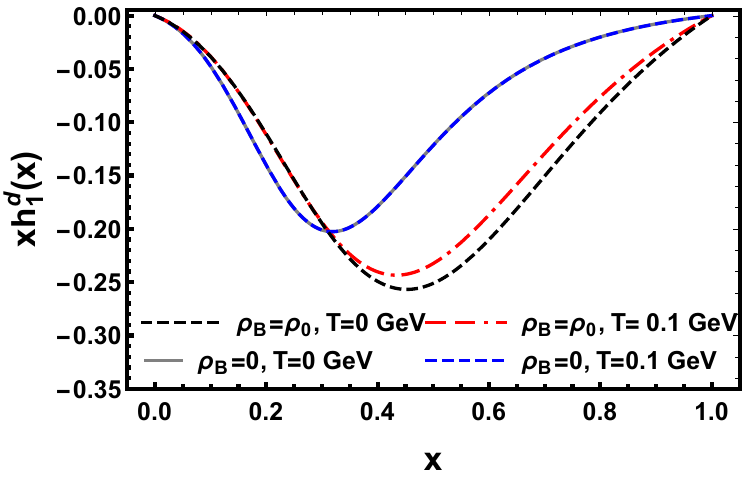}
        \hspace{0.03cm} 	 
    \end{minipage}
    \caption{\label{figT} The leading-twist proton parton distribution functions as a function of the longitudinal momentum fraction $x$ for fixed values of temperature $T=0$ and 0.1 (GeV) at baryon densities $\rho_B=0$ and $\rho_B=\rho_0$ in isospin symmetric nuclear matter, i.e., $\eta=0$. The top (bottom) row corresponds to the $u$ ($d$) quark distribution, while the first, second, and third columns represent the unpolarized distribution $xf_1(x)$, helicity distribution $xg_1(x)$, and transversity distribution $xh_1(x)$, respectively.}
\end{figure*}

To examine the interplay between baryon density and temperature $T$ (GeV), we present the corresponding quark distributions in Fig. \ref{figT} as a function of longitudinal momentum fraction $x$ in symmetric nuclear matter for two different values of baryon density. The PDFs are essentially insensitive to temperature at zero baryon density, whereas finite density effects become increasingly temperature dependent at saturation density. Increasing $T$ (GeV) from 0 to 0.1 partly counteracts the reduction of the effective masses induced by finite baryon density, leading to a slightly enhanced peak and a displacement of the peak toward lower $x$. This behavior is consistent for $xf_1^u$, $xf_1^d$, $xg_1^u$ and $xh_1^u$.
Quantitatively, Figs. \ref{figT}(a) and \ref{figT}(d) show that the amplitude of the distributions $xf_1^u$ and $xf_1^d$ shift from 1.892 at $x=0.53$ to 1.897 at $x=0.51$ and 0.581 at $x=0.35$ to 0.589 at $x=0.34$, respectively. Similarly, the peak of $xg_1^u$ distribution increases from 1.304 at $x=0.56$ to 1.308 at $x=0.54$ for the given values of $T=0$ and 0.1 GeV, respectively, as shown in Fig. \ref{figT}(b). For the case of transversity distribution (see Fig. \ref{figT}(c)), the peak value shifts from 1.176 at $x=0.54$ to 1.201 at $x=0.52$ when the temperature rises from $T=0$ to 0.1 GeV.
In contrast, the $xg_1^d$ distribution exhibits a nonuniform response to increasing temperature at the saturation density (see Fig. \ref{figT}(e)). While the negative contribution in the low-$x$ region increases with a shift of 0.019 value and the zero-cross over point moves toward higher $x$, the magnitude of the distribution decreases in the high-$x$ region from 0.145 to 0.124. Thus, for the case of $xg_1^d$, the combined effects of baryon density and temperature modify not only the magnitude and peak position but also the shape of the distribution.
Figure \ref{figT}(f) shows the case of $xh_1^d$ in which the peak magnitude of the distribution is found to decrease from 0.257 to 0.243 as the temperature increases to $T=0.1$ GeV from zero. 
The effects observed in the PDFs are transported to their corresponding momentum fractions carried by quarks, axial and tensor charges, as summarized in Tables \ref{tabfq}, \ref{tabAC}, and \ref{tabTC}, respectively.

\subsection{Isospin asymmetry effects on parton distribution functions}
\begin{figure*}
    \centering
    \begin{minipage}[c]{0.98\textwidth}
        (a)\includegraphics[width=4.7cm]{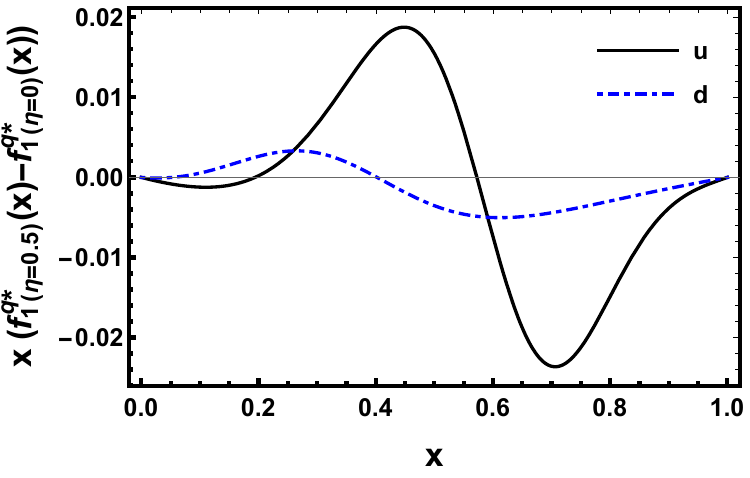}
        \hspace{0.03cm}
        (b)\includegraphics[width=4.7cm]{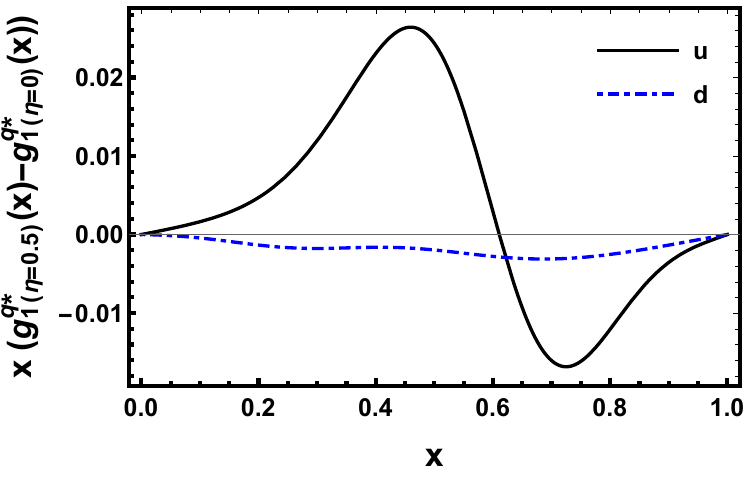}
        \hspace{0.03cm}	 
        (c)\includegraphics[width=4.7cm]{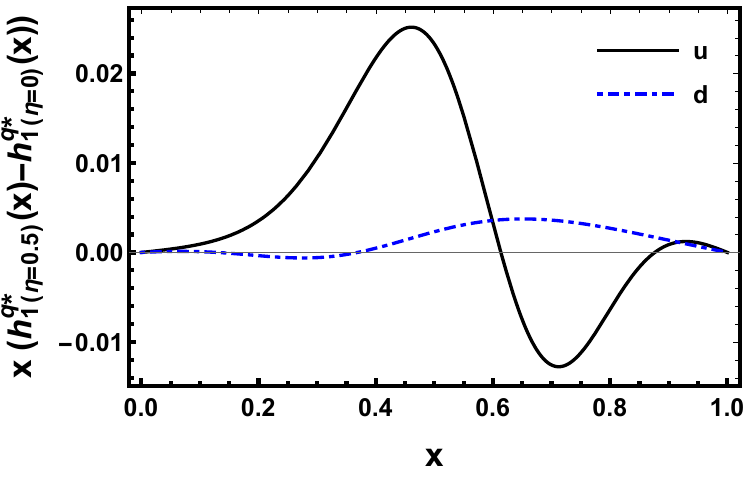}
        \hspace{0.03cm}
    \end{minipage}
    \caption{\label{figI3} Difference between the (a) unpolarized, (b) helicity, and (c) transversity quark distributions in asymmetric ($\eta=0.5$) and symmetric ($\eta=0$) nuclear matter for the proton at saturation density $\rho_B=\rho_0$ and zero temperature.}
\end{figure*}

To assess the impact of baryon density and isospin asymmetry $\eta$ on the quark distributions, we present the difference of PDFs between pure neutron matter ($\eta=0.5)$ and symmetric nuclear matter ($\eta=0)$ in Fig. \ref{figI3} as a function of longitudinal momentum fraction $x$ at zero temperature and saturation density. The difference of unpolarized PDFs for $u$ quark exhibits a characteristic sign change, being negative at small $x$, positive at intermediate $x$, and again negative at large $x$. The zero-cross over lies at $x=0.19$ and 0.59, which is shown in Fig. \ref{figI3}(a). 
For the case of $d$ quark, the distribution shows a positive change in the range $x<0.4$ and then flips to negative.
This indicates that $\eta$ enhances the intermediate-$x$ contribution while suppressing the small- and large-$x$ contribution for the case of unpolarized PDF. Unlike the unpolarized distributions, the helicity distributions exhibit a distinct flavor dependence as the $u$ quark contribution undergoes a reversal of sign at $x=0.62$, whereas the $d$ quark contribution remains negative over the whole range (see Fig. \ref{figI3}(b)). The transversity distributions presented in Fig. \ref{figI3}(c) display another distinct flavor dependence, with both $u$ and $d$ quark contributions exhibiting sign changes at $x=0.61$ and 0.42, respectively, but in opposite directions. In all three cases, the $u$ quark distribution shows a substantially stronger response to isospin asymmetry than the $d$ quark distribution. 
The stronger response of $u$ quark follows the mass shifts of Section \ref{Sec2}. At $\rho_0$ and $T=0$ GeV, $m_u^\ast$ rises by 2.69$\%$, while the $m_d^\ast$ falls by 0.17$\%$ (Table \ref{tab:dmass}).
The values of momentum fractions carried by quarks, axial and tensor charge as a function of $\eta$ are presented in Table \ref{tabfq}, \ref{tabAC} and \ref{tabTC}, respectively, for zero and finite $T=0.1$ GeV. The comparison indicates that the modifications induced by isospin asymmetry are comparatively smaller than those arising from variations in temperature $T$ (GeV) and baryon density $\rho_B$. 

\begin{table}[ht]
    \centering
    \renewcommand{\arraystretch}{1.3}
    \begin{tabular}{|c|c|c|c|c|c|c|c|c|}
        \hline
        
        \multirow{3}{*}{baryon density} &
        \multicolumn{4}{c|}{$\langle x \rangle^u$} &
        \multicolumn{4}{c|}{$\langle x \rangle^d$} \\ \cline{2-9}
        
        & \multicolumn{2}{c|}{$T=0$ GeV} & \multicolumn{2}{c|}{$T=0.1$ GeV} & \multicolumn{2}{c|}{$T=0$ GeV} & \multicolumn{2}{c|}{$T=0.1$ GeV} \\ \cline{2-9}
        
        & $~~\eta=0~~$ & $\eta=0.5$ & $~~\eta=0~~$ & $\eta=0.5$ & $~~\eta=0~~$ & $\eta=0.5$ & $~~\eta=0~~$ & $\eta=0.5$ \\
        
        \hline
        
        0 & 0.688 & 0.688 & 0.688 & 0.688 & 0.238 & 0.238 & 0.238 & 0.238 \\ \hline
        0.25$\rho_0$ & 0.720 & 0.720 & 0.715 & 0.715 & 0.250 & 0.249 & 0.248 & 0.247\\ \hline
        0.5$\rho_0$ & 0.750 & 0.750 & 0.741 & 0.742 & 0.262 & 0.259 & 0.258 & 0.256 \\ \hline
        0.75$\rho_0$ & 0.780 & 0.779 & 0.766 & 0.767 & 0.273 & 0.270 & 0.268 & 0.265 \\ \hline
        $\rho_0$ & 0.806 & 0.804 & 0.789 & 0.790 & 0.284 & 0.279 & 0.277 & 0.274\\ \hline
        
    \end{tabular}
    \caption{Second Mellin moment of unpolarized parton distribution function $\langle x \rangle^q$ for different baryon density $\rho_B$ of the hot and isospin-asymmetric nuclear matter at discrete values of temperature $T$ (GeV) and isospin-asymmetry $\eta$ for the $u$ and $d$ quark flavors of the proton.}
    \label{tabfq}
\end{table}
\begin{table}[ht]
    \centering
    \renewcommand{\arraystretch}{1.3}
    \begin{tabular}{|c|c|c|c|c|c|c|c|c|}
        \hline
        
        \multirow{3}{*}{baryon density} &
        \multicolumn{4}{c|}{$\Delta u$} &
        \multicolumn{4}{c|}{$\Delta d$} \\ \cline{2-9}
        
        & \multicolumn{2}{c|}{$T=0$ GeV} & \multicolumn{2}{c|}{$T=0.1$ GeV} & \multicolumn{2}{c|}{$T=0$ GeV} & \multicolumn{2}{c|}{$T=0.1$ GeV} \\ \cline{2-9}
        
        & $~~\eta=0~~$ & $\eta=0.5$ & $~~\eta=0~~$ & $\eta=0.5$ & $~~\eta=0~~$ & $\eta=0.5$ & $~~\eta=0~~$ & $\eta=0.5$ \\
        
        \hline
        
        0 & 1.095 & 1.095 & 1.095 & 1.095 & $-0.183$ & $-0.183$ & $-0.183$ & $-0.183$ \\ \hline
        0.25$\rho_0$ & 1.050 & 1.052 & 1.056 & 1.058 & $-0.126$ & $-0.127$ & $-0.136$ & $-0.135$\\ \hline
        0.5$\rho_0$ & 1.006 & 1.012 & 1.019 & 1.023 & $-0.062$ & $-0.062$ & $-0.083$ & $-0.081$ \\ \hline
        0.75$\rho_0$ & 0.964 & 0.974 & 0.984 & 0.990 & $0.008$ & 0.006 & $-0.026$ & $-0.024$\\ \hline
        $\rho_0$ & 0.923 & 0.938 & 0.950 & 0.959 & $0.079$ & 0.074 & $0.032$ & 0.034 \\ \hline
        
    \end{tabular}
    \caption{Valence quark contributions of the axial charge $\Delta q$ for different baryon density $\rho_B$ of the hot and isospin-asymmetric nuclear matter at discrete values of temperature $T$ (GeV) and isospin-asymmetry $\eta$ for the $u$ and $d$ quark flavors of the proton.}
    \label{tabAC}
\end{table}
\begin{table}[ht]
    \centering
    \renewcommand{\arraystretch}{1.3}
    \begin{tabular}{|c|c|c|c|c|c|c|c|c|}
        \hline
        
        \multirow{3}{*}{baryon density} &
        \multicolumn{4}{c|}{$\delta u$} &
        \multicolumn{4}{c|}{$\delta d$} \\ \cline{2-9}
        
        & \multicolumn{2}{c|}{$T=0$ GeV} & \multicolumn{2}{c|}{$T=0.1$ GeV} & \multicolumn{2}{c|}{$T=0$ GeV} & \multicolumn{2}{c|}{$T=0.1$ GeV} \\ \cline{2-9}
        
        & $~~\eta=0~~$ & $\eta=0.5$ & $~~\eta=0~~$ & $\eta=0.5$ & $~~\eta=0~~$ & $\eta=0.5$ & $~~\eta=0~~$ & $\eta=0.5$ \\
        
        \hline
        
        0 & 1.113 & 1.113 & 1.113 & 1.113 & $-0.287$ & $-0.287$ & $-0.287$ & $-0.287$ \\ \hline
        0.25$\rho_0$ & 1.051 & 1.053 & 1.060 & 1.061 & $-0.296$ & $-0.295$ & $-0.295$ & $-0.294$ \\ \hline
        0.5$\rho_0$ & 0.988 & 0.993 & 1.007 & 1.010 & $-0.312$ & $-0.309$ & $-0.306$ & $-0.305$ \\ \hline
        0.75$\rho_0$& 0.926 & 0.935 & 0.955 & 0.960 & $-0.332$ & $-0.328$ & $-0.322$ & $-0.319$ \\ \hline
        $\rho_0$ & 0.866 & 0.881 & 0.906 & 0.913 & $-0.356$ & $-0.349$ & $-0.340$ & $-0.336$ \\ \hline
        
    \end{tabular}
    \caption{Valence quark contributions of the tensor charge $\delta q$ for different baryon density $\rho_B$ of the hot and isospin-asymmetric nuclear matter at discrete values of temperature $T$ (GeV) and isospin-asymmetry $\eta$ for the $u$ and $d$ quark flavors of the proton.}
    \label{tabTC}
\end{table}
\section{Parton distribution functions in nuclei \label{Sec:finite}}
In this section, we compute the parton distributions of proton bound in various nuclei, including $^{40}\mathrm{Ca}$, $^{56}\mathrm{Fe}$, $^{197}\mathrm{Au}$, and $^{208}\mathrm{Pb}$. To approximate parton distributions in these nuclei, we employ a Woods-Saxon-type density distribution of nucleons, which is expressed as follows \cite{Chamon:2002},

\begin{equation}
    \rho_i(r) = \frac{\rho_{i,0}}{1 + \exp \left[ \frac{r - R_i}{a_i} \right]},
\end{equation}
where $\rho_{i,0}$ ($i=p,n$) is not a free parameter but is instead determined by the normalization condition, $4\pi \int_{0}^{\infty} \rho_i(r) r^2 dr = Z$ for proton and $N$ for neutron.  
The surface behavior of the density profile is governed by two factors: the diffuseness parameter $a_i$ and the half-density radius $R_i$, both in fm. For protons and neutrons, we take these parameters from the fits of Ref. \cite{Chamon:2002} to Dirac-Hartree-Bogoliubov densities, as in Ref. \cite{Ma:2024}.
\begin{equation}
    R_{{p}} = 1.81 Z^{1/3} - 1.12, \quad R_{{n}} = 1.49 N^{1/3} - 0.79,
\end{equation}
and
\begin{equation}
    a_{\text{p}} = 0.47 - 0.00083 Z, \quad a_{\text{n}} = 0.47 + 0.00046 N.
\end{equation}
	In order to obtain a medium impact on parton distribution within finite nuclei, we utilize the CQMF model to solve the coupled equations of motion governing the mesonic fields, which are coupled to the baryonic density of nucleons. The total baryonic density of the nucleus $\rho_B(r)$ is then given by the sum over proton and neutron contributions, $\rho_B(r) = \sum\limits_{i=p,n} \rho_i(r)$, where the nuclei are treated in the local density approximation. From these densities, we obtain the in-medium masses of quarks, diquarks, and protons confined within various nuclei, specifically $^{40}\mathrm{Ca}$, $^{56}\mathrm{Fe}$, $^{197}\mathrm{Au}$, and $^{208}\mathrm{Pb}$. 
\begin{figure}
    \centering
    \includegraphics[width=0.5\linewidth]{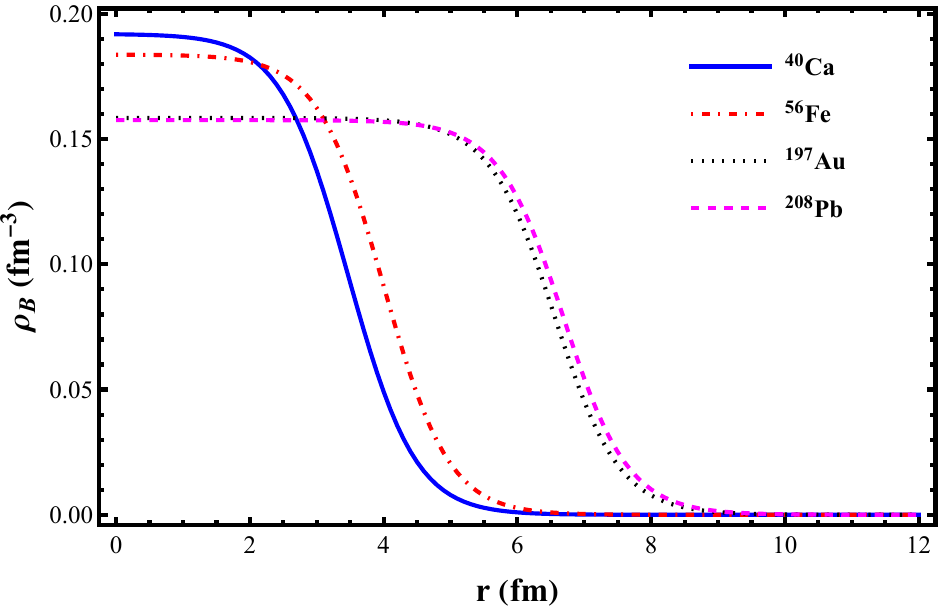}
    \caption{The radial profile of total baryonic density $\rho_B(r)$ in $\mathrm{fm}^{-3}$ for various nuclei, including $^{40}\mathrm{Ca}$, $^{56}\mathrm{Fe}$, $^{197}\mathrm{Au}$, and $^{208}\mathrm{Pb}$, given by the Woods-Saxon distributions of the proton and neutron densities.}
    \label{fig_density_r}
\end{figure}
\begin{figure*}
    \centering
    \begin{minipage}[c]{0.98\textwidth}
        (a)\includegraphics[width=4.7cm]{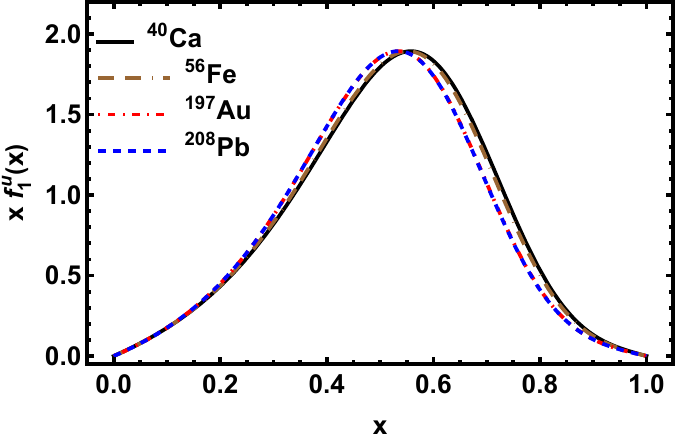}
        \hspace{0.03cm}
        (b)\includegraphics[width=4.7cm]{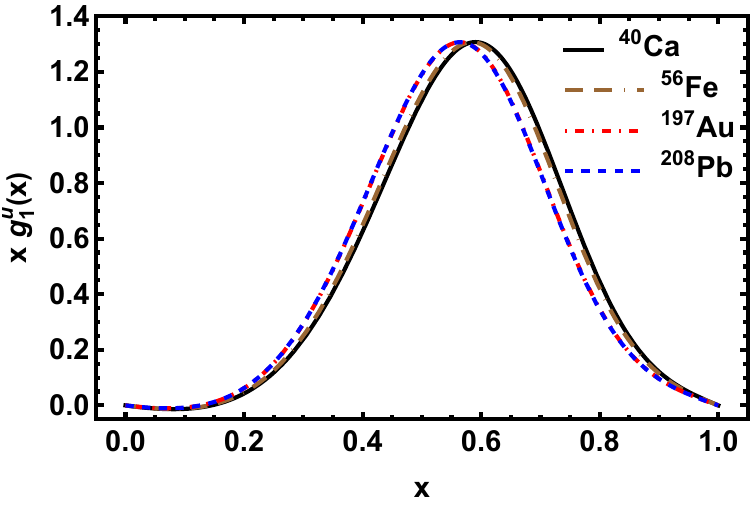}
        \hspace{0.03cm}	 
        (c)\includegraphics[width=4.7cm]{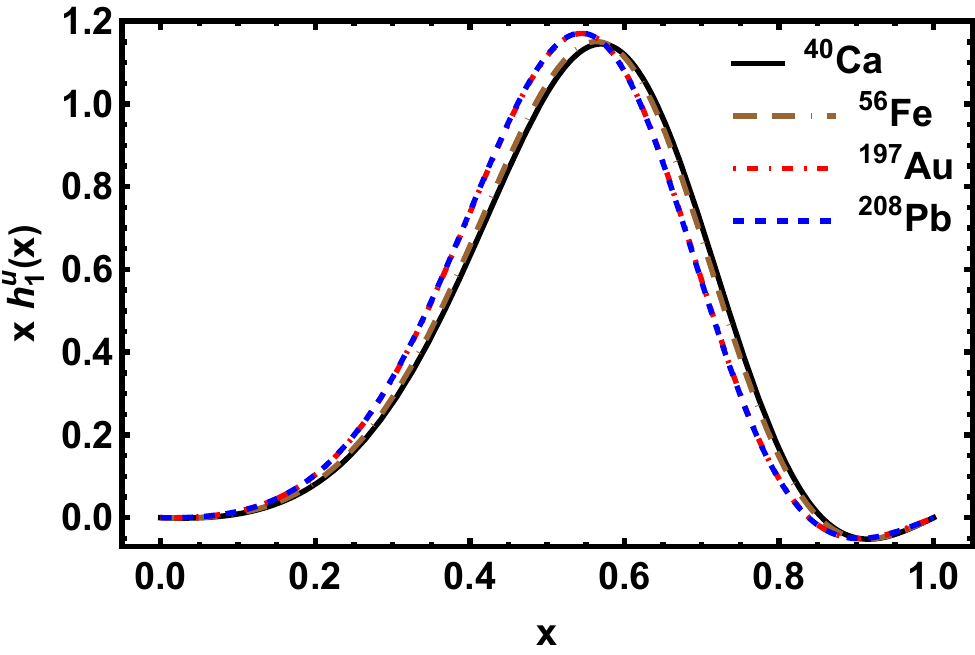}
        \hspace{0.03cm}
        (d)\includegraphics[width=4.7cm]{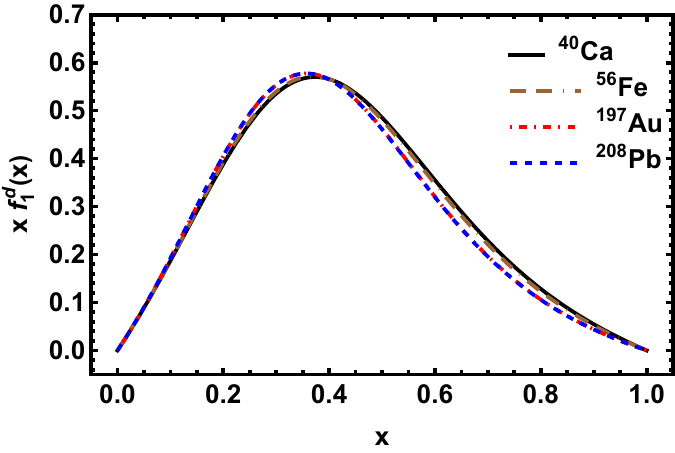}
        \hspace{0.03cm}	
        (e)\includegraphics[width=4.7cm]{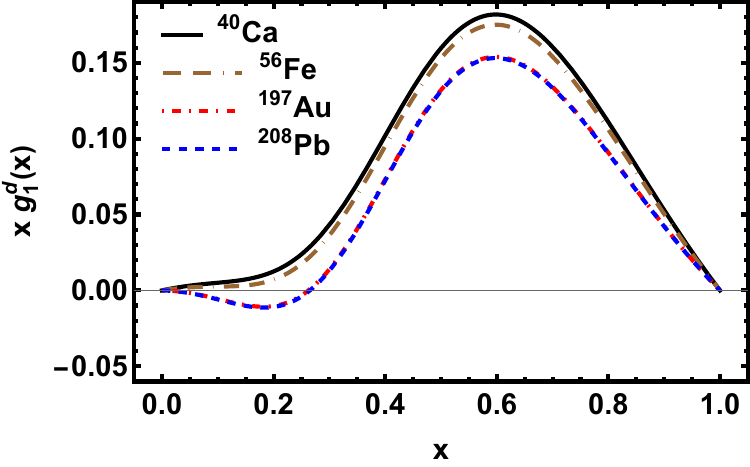}
        \hspace{0.03cm}
        (f)\includegraphics[width=4.7cm]{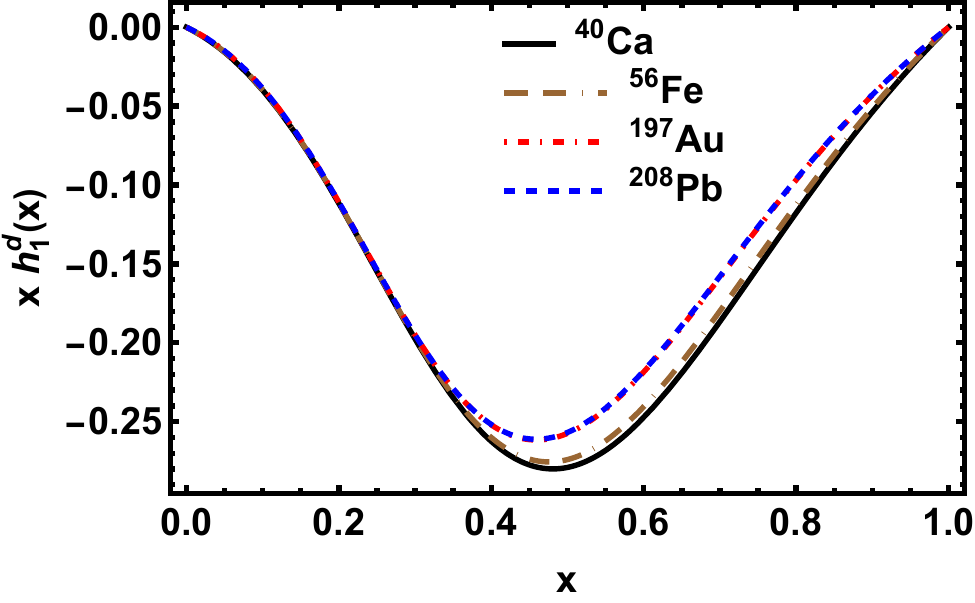}
        \hspace{0.03cm}
    \end{minipage}
    \caption{The leading-twist proton parton distribution functions are depicted as a function of the longitudinal momentum fraction $x$ for various nuclei: $^{40}\mathrm{Ca}$, $^{56}\mathrm{Fe}$, $^{197}\mathrm{Au}$, and $^{208}\mathrm{Pb}$. The top row illustrates the $u$ quark distributions, while the bottom row shows the $d$ quark distributions. The columns are organized to display the unpolarized distribution $xf_1(x)$ in subfigures (a) and (d), the helicity distribution $xg_1(x)$ in subfigures (b) and (e), and the transversity distribution $xh_1(x)$ in subfigures (c) and (f).}
    
    \label{fig_pdf_nuclei}
\end{figure*} 
\begin{figure*}
    \centering
    \begin{minipage}[c]{0.98\textwidth}
        (a)\includegraphics[width=4.7cm]{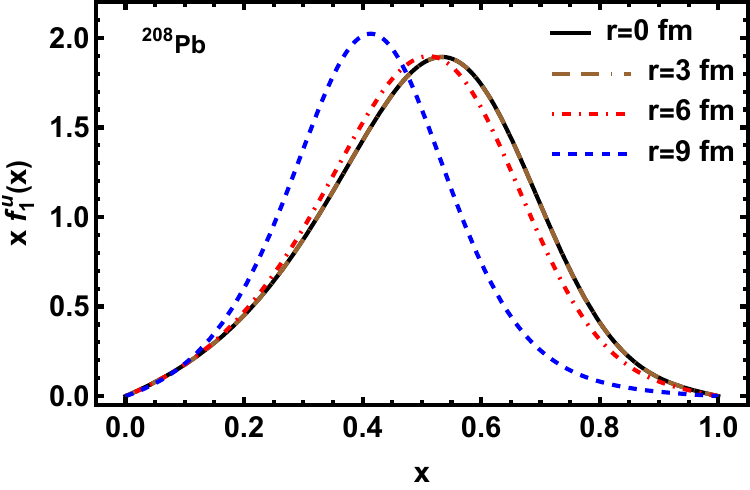}
        \hspace{0.03cm}
        (b)\includegraphics[width=4.7cm]{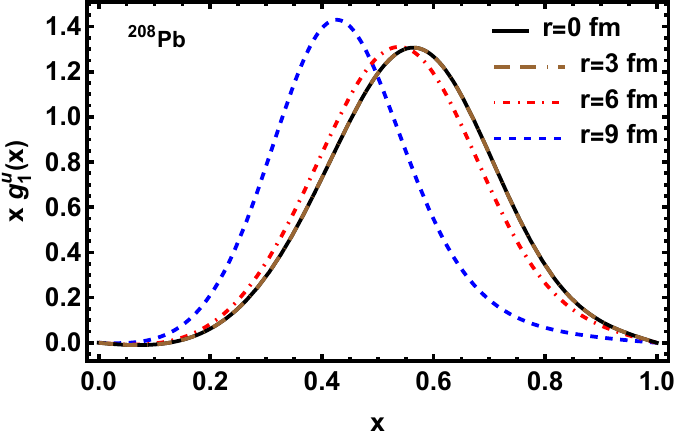}
        \hspace{0.03cm}	 
        (c)\includegraphics[width=4.7cm]{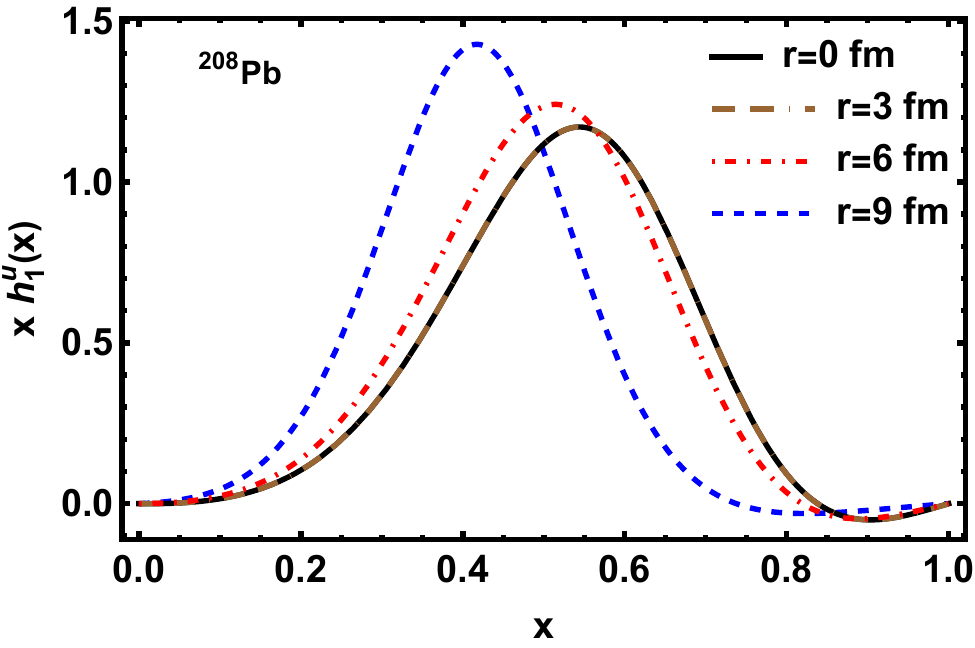}
        \hspace{0.03cm}
        (d)\includegraphics[width=4.7cm]{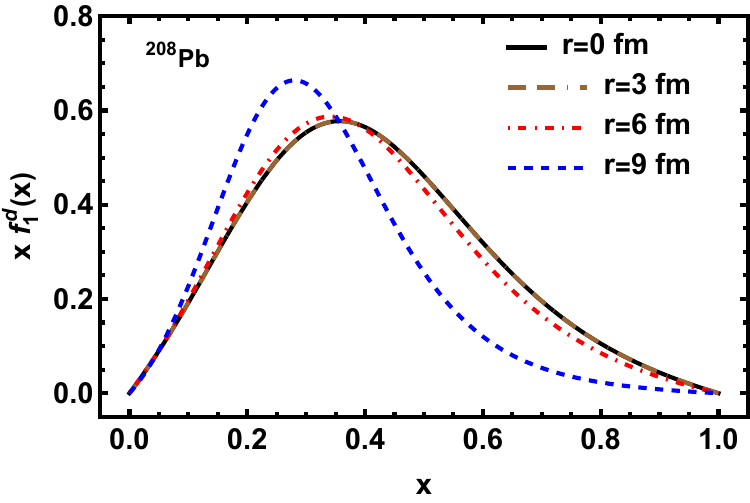}
        \hspace{0.03cm}	
        (e)\includegraphics[width=4.7cm]{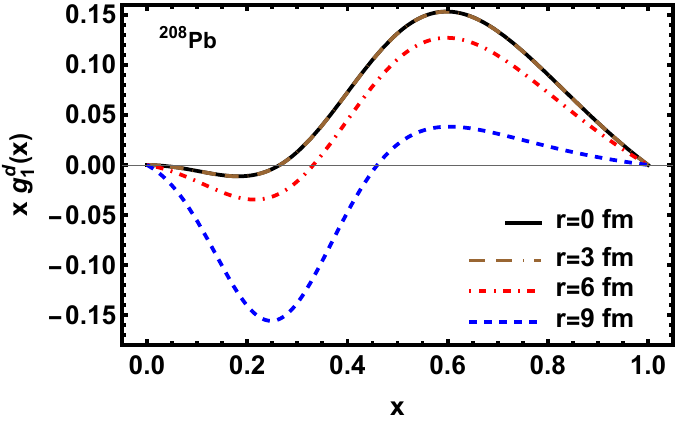}
        \hspace{0.03cm}
        (f)\includegraphics[width=4.7cm]{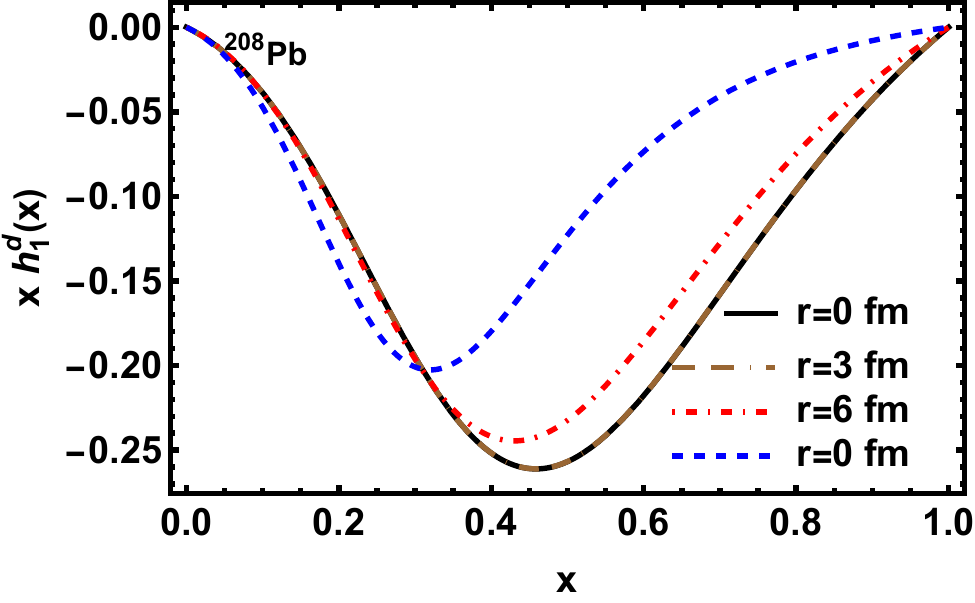}
        \hspace{0.03cm}
    \end{minipage}
    \caption{The leading-twist proton parton distribution functions are depicted as a function of the longitudinal momentum fraction $x$ for $^{208}\mathrm{Pb}$ nucleus at various values of radial distance $r$. The top row illustrates the $u$ quark distributions, while the bottom row shows the $d$ quark distributions. The columns are organized to display the unpolarized distribution $xf_1(x)$ in subfigures (a) and (d), the helicity distribution $xg_1(x)$ in subfigures (b) and (e), and the transversity distribution $xh_1(x)$ in subfigures (c) and (f).}
    \label{fig_pdf_nuclei1}
\end{figure*} 
We now plot the total baryonic density of the nuclei, $\rho_B(r)$, varying with the radial distance $r$ from the center of the nuclei $^{40}\mathrm{Ca}$, $^{56}\mathrm{Fe}$, $^{197}\mathrm{Au}$, and $^{208}\mathrm{Pb}$ in Fig.~\ref{fig_density_r}. The total baryonic density decreases with increasing distance from the center of the nucleus. Specifically, the values of $\rho_B$ in the center of the nuclei ($r=0$) are found to be 0.191, 0.184, 0.158, and 0.157 $\mathrm{fm}^{-3}$ for $^{40}\mathrm{Ca}$, $^{56}\mathrm{Fe}$, $^{197}\mathrm{Au}$, and $^{208}\mathrm{Pb}$, respectively. For these values of $\rho_B$, the leading-twist proton distributions with respect to the longitudinal momentum fraction $x$ for $^{40}\mathrm{Ca}$, $^{56}\mathrm{Fe}$, $^{197}\mathrm{Au}$, and $^{208}\mathrm{Pb}$ nuclei are plotted in Fig.~\ref{fig_pdf_nuclei}. The results of unpolarized PDFs for $u$ and $d$ quarks for different nuclei are presented in  Figs.~\ref{fig_pdf_nuclei}(a) and~\ref{fig_pdf_nuclei}(d). For light nuclei, the peak position of $f^{u}_1$ is centered around $x \approx 0.55$, while the peak position of $f^{d}_1$ lies near $x \approx 0.35$. Going from light to heavy nuclei, both peaks shift toward lower $x$, reflecting how these PDFs respond to the baryonic density of the nuclear medium. 
In Figs.~\ref{fig_pdf_nuclei}(b) and~\ref{fig_pdf_nuclei}(e), we plot the results of the helicity distribution of the light quarks, $g^{u}_1$ and $g^{d}_1$, as a function of $x$ for all nuclei considered in this study. As the central density falls, the valence $u$ quark distribution develops a modest rise near its maximum. In contrast, the $d$ quark distribution is progressively suppressed for $x > 0.27$ for heavy nuclei. Below this value of $x$, a negative modification appears only for the heavy nuclei as illustrated in Fig.~\ref{fig_pdf_nuclei}(e). Similarly, the results of the transversity distribution $h_1(x)$ in Figs.~\ref{fig_pdf_nuclei}(c) and~\ref{fig_pdf_nuclei}(f) are displayed. We observe a positive contribution of $h^u_1(x)$ till $x < 0.85$, but its effect turns negative beyond this point. On the other hand, $h^d_1(x)$ contributes negatively across the considered $x$ range.
Further, we present these leading-twist proton parton distributions for the $^{208}\mathrm{Pb}$ nucleus at various values of radial distance $r=0$, 3, 6, 9~fm in Fig.~\ref{fig_pdf_nuclei1}. The distributions at $r=0$ and 3 fm coincide, since the density stays close to its central value there (Fig. \ref{fig_density_r}). At $r=6$ fm, in the nuclear surface, the density is lower and the distributions move part of the way back toward their vacuum forms, while at $r=9$ fm, where the density has almost vanished, they coincide with the vacuum distributions. At the center of each nucleus, the axial charge $g_A=\Delta u - \Delta d$ of the bound proton is 0.699 for $^{40}\mathrm{Ca}$, 0.741 for $^{56}\mathrm{Fe}$, 0.815 for $^{197}\mathrm{Au}$, and 0.817 for $^{208}\mathrm{Pb}$, against 1.28 in vacuum, so it increases as the central density decreases. 
In Ref. \cite{Cloet:2006bq}, the axial charge of the bound proton tends to decrease with mass number, from $1.267$ for the free proton to $1.19$ in $^{7}$Li, $1.16$ in $^{11}$B, and $1.15$ in $^{15}$N and $^{27}$Al, and it reaches $1.05$ in nuclear matter. The calculation of Ref. \cite{Cloet:2006bq} describes the nucleus in a relativistic shell model with Woods-Saxon scalar and vector potentials of the same depth for every nucleus. The spin-dependent distributions there are carried mainly by the valence proton or proton hole, which feels the mean fields averaged over its single-particle state. The values obtained here are instead taken at the center of each nucleus, where the central density falls from $^{40}$Ca to $^{208}$Pb, so the opposite dependence on mass number may reflect the different quantity computed. In both calculations, the axial charge is lower in nuclear matter than in the free proton.

\section{Summary \label{summary}}  
The impact of hot and isospin-asymmetric nuclear matter and finite nuclei on the proton parton distribution functions (PDFs) is investigated within the diquark spectator model. The medium effects are incorporated through the effective masses of the quarks, diquarks, and proton, which are calculated within the chiral SU(3) quark mean-field model. The resulting modifications of these effective masses arise from their interactions with the scalar fields and consequently influence the light-front wave functions and the partonic structure of the proton. The effective masses decrease with increasing baryon density. At temperature $T=0.1$ GeV, they lie above their $T=0$ values, while in pure neutron matter, $m_u^*$ lies above and $m_d^*$ below its value in symmetric nuclear matter, by at most $2.7\%$.

The in-medium unpolarized $f_1^q$, helicity $g_1^q$, and transversity $h_1^q$ distributions for both quark flavors are calculated as functions of longitudinal momentum fraction $x$ at different baryon densities $\rho_B$, temperatures $T$ (GeV), and isospin asymmetries $\eta$. 
The impact of the medium conditions on the parton distributions follows the order: baryon density $>$ temperature $>$ isospin asymmetry. 
As a function of baryon density, the distributions of $xf_1^u$, $xf_1^d$, $xg_1^u$, and $xh_1^u$ exhibit a decrease in their peak values, accompanied by a shift of the peaks toward higher $x$ values. In contrast, the distribution $xg_1^d$ exhibits a decrease in the magnitude of its negative contribution in the low-$x$ region, accompanied by a shift of the crossover point toward lower $x$. Conversely, the positive contribution increases in magnitude in the  high-$x$ domain when the medium becomes denser. For the case of $xh_1^d$, the distribution exhibits an increase in the magnitude of its negative contribution in the whole span of $x$ region when the baryon density increases from zero to its saturation value.
Corresponding to the unpolarized, helicity, and transversity distributions, their Mellin moments are also computed. Second Mellin moment of the $f_1^q$ increases with increasing the density of the matter, whereas the axial charge $g_A$ and the $u$ quark tensor charge decrease and the $d$ quark tensor charge grows in magnitude. The axial charge of the proton falls from $1.28$ in vacuum to $0.844$ at saturation density. This agrees in sign with the NJL quark-diquark calculation \cite{Cloet:2006bq}, where the axial charge of the bound proton falls from $1.267$ for the free proton to $1.05$ in nuclear matter, while the reduction obtained here is larger.  A comparison of the spin-dependent and spin-independent distributions reveals more pronounced changes in the spin-dependent distributions.

At zero baryon density, the distributions remain essentially unchanged with temperature and isospin asymmetry. However, at saturation density, the distributions vary with $T$ and $\eta$. Moving from $T=0$ to $0.1$ GeV, the distributions $xf_1^u$, $xf_1^d$, $xg_1^u$, and $xh_1^u$ show a slight increase in their peak heights, and their peaks shift toward smaller $x$. In contrast, the negative peak of $xh_1^d$ becomes shallower, while also moving to smaller $x$. For $xg_1^d$, the negative contribution increases in the low-$x$ region, whereas the positive contribution decreases in the high-$x$ region. To study the impact of isospin asymmetry, the difference between asymmetric ($\eta=0.5$) and symmetric ($\eta=0$) nuclear matter over the distributions is presented. The $u$ quark distributions exhibit a characteristic sign change over the range of $x$ for all three cases: unpolarized, helicity, and transversity. In contrast, the $d$ quark helicity distribution remains negative, while the other $d$ quark distributions exhibit opposite signs in the low- and high-$x$ regions. In the corresponding Mellin moments, the effect of $\eta$ is small, while at saturation density, raising $T$ to 0.1 GeV increases the axial charge from 0.844 to 0.918.

In addition, we also examined parton distributions of a proton bound in various nuclei, including $^{40}\mathrm{Ca}$, $^{56}\mathrm{Fe}$, $^{197}\mathrm{Au}$, and $^{208}\mathrm{Pb}$ at zero temperature by employing Woods-Saxon-type density distributions. 
As the central density falls from $^{40}$Ca to $^{208}$Pb, the peaks of $xf_1^q$ move to lower $x$, $xg_1^u$ rises slightly near its peak, the positive part of $xg_1^d$ is suppressed, and $xh_1^d$ stays negative for all $x$. At the center of each nucleus, the axial charge of the proton rises from $0.699$ in $^{40}$Ca to $0.817$ in $^{208}$Pb as the central baryon density decreases, which suggests that the modification at the center follows the central density rather than the mass number. In Ref. \cite{Cloet:2006bq}, the axial charge of the bound proton instead decreases with mass number, and the spin-dependent distributions are carried mainly by the valence proton or proton hole, which feels the mean fields averaged over its single-particle state. The opposite trend may therefore reflect the different quantity computed, since the values here are taken at the center of the nucleus. In the present work, we studied the in-medium quark dynamics by incorporating effective masses through scalar-field interactions. Future extensions will include vector-field interactions to provide a more comprehensive description of the nuclear medium and examine their effects on the quark dynamics and the resulting parton distribution functions.

\section{Acknowledgements}
H.D. would like to thank the Science and Engineering Research Board, Anusandhan-National Research Foundation, Government of India under the scheme SERB-POWER Fellowship (Ref No. SPF/2023/000116) for financial support.
\appendix

\section{Derivatives of the effective nucleon mass}
\label{app:gN}

Here we derive Eq. (\ref{eq:gclosed}), writing $\varphi$ for either of the two fields $\sigma$ and $\delta$ in Eq. (\ref{eq:gN}), so that both couplings take the form $g_\varphi^i = -\partial M_i^*/\partial\varphi$. The spin correlation energy $E_{\rm spin}^i$ and the confinement strength $k_c$ are held fixed, so the fields reach $M_i^*$ only through the quark masses, along the chain $\varphi \to m_q^* \to (e_q, \langle p_q^2\rangle) \to (E_i, \langle P_{\rm c.m.}^2\rangle_i) \to M_i^*$. By Eqs. (\ref{eq:eigen}) and (\ref{eq:p2}), $e_q$ and $\langle p_q^2\rangle$ depend on $m_q^*$ alone, so the entire dependence on $\varphi$ enters through the derivatives of the quark masses. From Eq. (\ref{eq:mud}) these are
\begin{equation}
    \frac{\partial m_q^*}{\partial\sigma} = -\frac{g_s}{\sqrt{2}}, \qquad \frac{\partial m_q^*}{\partial\delta} = -\frac{g_s}{\sqrt{2}}\, I_3^q, \qquad q = u, d.
    \label{eq:dmdphi}
\end{equation}
Since $e_q - m_q^* > 0$, Eq. (\ref{eq:eigen}) is equivalent to $(e_q - m_q^*)^2(e_q + m_q^*) = 9k_c/2$. Differentiating this form with respect to $m_q^*$ and dividing by $e_q - m_q^*$ gives
\begin{equation}
    2\left(e_q + m_q^*\right)\left(r_q - 1\right) + \left(e_q - m_q^*\right)\left(r_q + 1\right) = 0, \qquad r_q \equiv \frac{de_q}{dm_q^*},
\end{equation}
so that
\begin{equation}
    r_q = \frac{e_q + 3m_q^*}{3e_q + m_q^*}.
    \label{eq:dedm}
\end{equation}
The quark energy then responds to the fields as $\partial e_q/\partial\varphi = r_q\,\partial m_q^*/\partial\varphi$.
The mean square momentum of Eq. (\ref{eq:p2}) depends on $m_q^*$ both directly and through $e_q$. Writing it as $\langle p_q^2\rangle = N_q/D_q$, with $N_q = (11e_q + m_q^*)(e_q^2 - m_q^{*2})$ and $D_q = 6(3e_q + m_q^*)$, and using $de_q/dm_q^* = r_q$, we have
\begin{equation}
    \frac{dN_q}{dm_q^*} = \left(11r_q + 1\right)\left(e_q^2 - m_q^{*2}\right) + 2\left(11e_q + m_q^*\right)\left(r_q e_q - m_q^*\right), \qquad \frac{dD_q}{dm_q^*} = 6\left(3r_q + 1\right).
\end{equation}
The quotient rule then gives
\begin{equation}
    s_q \equiv \frac{d\langle p_q^2\rangle}{dm_q^*} = \frac{1}{D_q}\left(\frac{dN_q}{dm_q^*} - \langle p_q^2\rangle\,\frac{dD_q}{dm_q^*}\right),
    \label{eq:dp2dm}
\end{equation}
and the mean square momentum responds to the fields as $\partial\langle p_q^2\rangle/\partial\varphi = s_q\,\partial m_q^*/\partial\varphi$.
Since $E_{\rm spin}^i$ is constant, Eqs. (\ref{eq:EN}) and (\ref{eq:MN}) give
\begin{equation}
    \frac{\partial E_i}{\partial\varphi} = \sum_{q=u,d} n_q^i r_q\,\frac{\partial m_q^*}{\partial\varphi}, \qquad \frac{\partial\langle P_{\rm c.m.}^2\rangle_i}{\partial\varphi} = \sum_{q=u,d} n_q^i s_q\,\frac{\partial m_q^*}{\partial\varphi}.
\end{equation}
Differentiating $M_i^{*2} = E_i^2 - \langle P_{\rm c.m.}^2\rangle_i$ and inserting these sums gives
\begin{equation}
    \frac{\partial M_i^*}{\partial\varphi} = \frac{1}{M_i^*}\left(E_i\frac{\partial E_i}{\partial\varphi} - \frac{1}{2}\frac{\partial\langle P_{\rm c.m.}^2\rangle_i}{\partial\varphi}\right) = \frac{1}{M_i^*}\sum_{q=u,d} n_q^i c_q^i\, \frac{\partial m_q^*}{\partial\varphi},
    \label{eq:dMdphi}
\end{equation}
with $c_q^i = E_i r_q - s_q/2$, as in Eq. (\ref{eq:gclosed}). The energy $E_i$ in $c_q^i$ is the full nucleon energy of Eq. (\ref{eq:EN}), including $E_{\rm spin}^i$. Inserting the quark mass derivatives of Eq. (\ref{eq:dmdphi}) into Eq. (\ref{eq:dMdphi}) and changing the sign gives
\begin{equation}
	\begin{aligned}
		g_\sigma^i &= -\frac{1}{M_i^*}\sum_{q=u,d} n_q^i c_q^i \left(-\frac{g_s}{\sqrt{2}}\right) = \frac{g_s}{\sqrt{2}\, M_i^*}\sum_{q=u,d} n_q^i c_q^i, \\
		g_\delta^i &= -\frac{1}{M_i^*}\sum_{q=u,d} n_q^i c_q^i \left(-\frac{g_s}{\sqrt{2}}\, I_3^q\right) = \frac{g_s}{\sqrt{2}\, M_i^*}\sum_{q=u,d} n_q^i I_3^q c_q^i,
	\end{aligned}
\end{equation}
which is Eq. (\ref{eq:gclosed}).

\end{document}